\documentclass[lettersize,journal]{IEEEtran}

\usepackage[font=footnotesize]{caption}
\usepackage{subcaption}

\usepackage{graphicx}
\usepackage{amsmath}
\usepackage{amssymb}

\usepackage{multirow}
\usepackage{colortbl}
\usepackage{array}
\usepackage{tabularx}
\usepackage{adjustbox}
\usepackage{arydshln}
\usepackage{hhline}

\usepackage{xcolor} 
\usepackage{soul}
\usepackage{textcomp}
\usepackage{gensymb}

\usepackage{verbatim}

\usepackage{float}
\usepackage{stfloats}
\usepackage{flushend}
\usepackage[bottom]{footmisc}
\usepackage{placeins} 

\usepackage[inline]{enumitem}

\setlist{nosep}

\usepackage{amsthm}

\usepackage{siunitx}
\usepackage{epstopdf}
\usepackage{tikz}
\usetikzlibrary{
  chains,positioning,shapes.symbols,arrows,shapes,shadows,trees,
  matrix,calc
}

\usepackage{standalone}

\usepackage{cite}
\usepackage{cleveref}

\usepackage{blkarray}
\usepackage{cuted, nccmath}
\usepackage{lipsum}
\usepackage{comment}
\usepackage{atbegshi}

\usepackage{algorithm}
\usepackage{algpseudocode}

\IEEEoverridecommandlockouts


\makeatletter
\newcommand{\multiline}[1]{%
  \begin{tabularx}{\dimexpr\linewidth-\ALG@thistlm}[t]{@{}X@{}}
    #1
  \end{tabularx}
}
\makeatother

 \renewcommand{\arraystretch}{1.5}

\algnewcommand{\AND}{\wedge}

\begin{document}

\bstctlcite{IEEEexample:BSTcontrol}

\title{Run-Length-Limited ISI-Mitigation (RLIM) Coding for Molecular Communication}

\author{Melih~Şahin, ~\IEEEmembership{Student Member,~IEEE},
		Ozgur~B.~Akan,~\IEEEmembership{Fellow,~IEEE}
		\thanks{ Melih Şahin and O. B. Akan are with the Internet of Everything (IoE) Group,	Department of Engineering, University of Cambridge, Cambridge CB3 0FA, U.K. (e-mail: \ \{ms3195, oba21\}@cam.ac.uk).}
  \thanks{O. B. Akan is also with the Center for neXt-generation Communications (CXC), Department of Electrical and Electronics Engineering, Koç University, Istanbul 34450, Türkiye (e-mail: \ \ akan@ku.edu.tr).}

		\thanks{This work was supported in part by the AXA Research Fund (AXA Chair for Internet of Everything at Koç University).}
}

\maketitle

\thispagestyle{empty}


\begin{abstract}
Inter-symbol interference (ISI) limits reliability in diffusion-based molecular communication (MC) channels. We propose RLIM, a family of run-length-limited (RLL) codes that form fixed-size codebooks by minimizing the total number of 1-bits, increasing the per-symbol molecule budget under standard power normalizations and thus improving reliability. We develop a provably optimal linear-time greedy decoder that is equivalent to Viterbi decoding under a deterministic last-wins tie-break and has lower computational complexity; empirically, it outperforms first-wins and random Viterbi variants on RLL baselines. Extensive binomial and particle-tracking simulations show that RLIM achieves lower bit error rate (BER) than classical RLL and other prominent coding schemes across a broad range of scenarios.
\end{abstract}


%
\IEEEpeerreviewmaketitle

\section{Introduction}
%
%
%
%

\IEEEPARstart{M}{o}lecular communication (MC) targets environments where electromagnetic signaling is ineffective or undesirable, for example short-range links in aqueous, cluttered, or biologically sensitive media. Typical applications include in-body sensing and actuation (e.g., coordinating drug-delivery carriers) and interfaces with engineered or natural cells via ligand–receptor chemistry; outside the body, MC is relevant in confined settings such as porous media and microfluidic testbeds \cite{ozgur_hoca_survey,MC_main,nakano_eckford_haraguchi_2013}. These contexts motivate robust coding and detection mechanisms that respect diffusive transport and strict time/power budgets while delivering reliable communication.

In a generic MC channel, as shown in Fig. 1, a transmitter and a receiver are surrounded by a fluidic environment that facilitates the diffusion-based movement of particles \cite{ozgur_hoca_survey,human}. This diffusive and seemingly-random movement of particles is characterized by Brownian Motion \cite{MC_main,nakano_eckford_haraguchi_2013}. In this model, the transmitter releases a number of information molecules at the start of each signal interval, which are then detected by the receiver if they come into its vicinity. The receiver decodes the information sent through the channel based on when, how many, and what types of information molecules it absorbed. MC causes a high degree of inter-symbol interference (ISI), a phenomenon where the gradual accumulation of information molecules, not yet absorbed by the receiver, impairs the receiver's ability to correctly decode the intended signal \cite{ISI}.

\begin{figure}[t]
\centering
\includegraphics[width=1\linewidth]{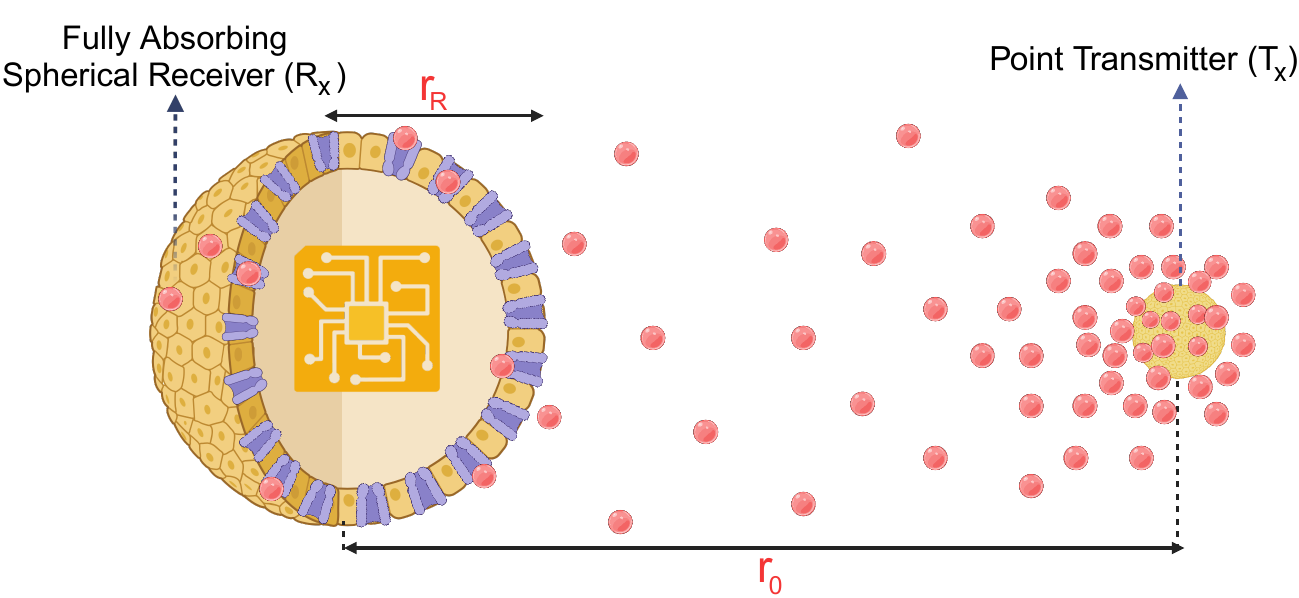}

\caption{MC Channel \cite{Sahin2024} }
\label{fig1}
\end{figure}

A number of coding schemes have been used or proposed in the MC literature, including ISI-mitigating codes \cite{best_channel_coding_2020}, ISI-free codes \cite{ISI-Free}, Uncoded BCSK \cite{BCSK}, and Hamming \cite{hamming,hamming2} codes. In terms of bit error rate (BER), ISI-mitigating codes have been shown to outperform these techniques \cite{best_channel_coding_2020}. In this work, we focus on constrained-sequence coding that explicitly regulates run-lengths to limit ISI. Run-length–limited (RLL) codes are classical in storage/recording \cite{RLL} and have been recently adapted to DNA coding \cite{DNA_RLL0,DNA_RLL1}; related constrained designs in MC include ISI-mitigating codes \cite{best_channel_coding_2020}. Decoding for such constrained codes is typically performed with Viterbi on a finite-state trellis, with practical variants differing mainly in how equal-metric ties are resolved \cite{mitViterbiNotes,original_viterbi,two_viterbi}.

We show that a last-wins Viterbi tie-break (equivalently, a proposed lower-complexity greedy correction) gives the lowest BER on our simulated MC channels, and we further select, within the admissible $(i,\infty)$–RLL set, a fixed-size subset that minimizes the total number of 1-bits which results in our proposed RLIM construction. Through extensive MC simulations across diverse scenarios, we demonstrate that our proposed RLIM codes achieve a significant BER advantage over prominent methods, including ISI-mitigating codes \cite{best_channel_coding_2020}.

The structure of this paper is as follows: In Section II, system model is provided. Section \ref{sec:generalisedcoding} gives the codebook constructions of the proposed RLIM codes. Error-correction and detection algorithms are given in Section \ref{sec:error} and \ref{sec:error_correct}. An analytical estimation of static threshold is available in Section \ref{sec:estimation}. In Section \ref{sec:performance}, MC channel simulation results are provided. This paper is concluded in Section V.

\section{System Model}
\label{sec:generalisedcoding}

One of the most  wide-spread MC modulation techniques \cite{modulation_techniques} is the binary concentration shift keying (BCSK) \cite{BCSK}. In BCSK, if a current signal interval corresponds to a 1-bit, a certain number of information molecules are released into the environment from the transmitter at the very start of that signal interval. If a current signal interval corresponds to a 0-bit, no information molecules are released. This paper uses BCSK modulation and assumes perfect time synchronization between the transmitter and receiver, both of which are common approaches in most MC coding studies \cite{ozgur_hoca_survey}.

The analytical function \cite{channel_characteristics} that gives the probability that an information molecule touches (i.e., gets absorbed by) the  fully absorbing spherical receiver until time $t$ (in seconds) is \begin{gather*}
F(t) = \frac{r_{R}}{r_{0}} \cdot \text{erfc}\!\left(\frac{r_{0} - r_{R}}{\sqrt{4 \cdot D \cdot t}}\right), \tag{1}
\end{gather*}where ${r_{R}}$ denotes the radius of the spherical receiver in $\si{\micro\metre}$, ${r_{0}}$ represents the distance between the center of the spherical receiver and the point transmitter in $\si{\micro\metre}$, $D$ is the diffusion coefficient of the fluidic environment in $\si{\micro\metre^2/\second}$, and $\text{erfc}(\cdot)$ denotes the complementary error function. The $i^{th}$ channel coefficient, ${p_{i}}$, is defined to be the probability that a molecule, emitted in the $k^{th}$ signal interval, gets absorbed during the $(k+i-1)$$^{th}$ signal interval. Channel coefficients \cite{ozgur_hoca_survey} are

\begin{gather*} p_{i} = F(i \cdot t_{s}) - F((i-1)\cdot t_{s}), \hskip 1em i =1,2,\ldots,L,\tag{2}\\[-2em]\end{gather*} where $L$ is the channel memory, and ${t_{s}}$ is the signal interval duration. Let $b_j$ represent the bit information transmitted in the ${(j-1)^{th}}$ previous signal slot (with $b_1$ indicating the bit from the current signal interval and $b_2$ indicating the bit from the previous interval). Through basic probability theory \cite{gaussian}, the random count, $N_{exp}$, to be detected in the current signal interval can then be modeled by using a summation of binomial distribution expressed as

\begin{equation}
    N_{exp} \sim \Big( \sum_{j=1}^{L} b_j \cdot \mathcal{B}(M,p_j)\Big) + \mathcal{N}(0,\sigma_n^2)
    \tag{3},
\end{equation}where $M$ is the number of  information molecules emitted at the start of each signal interval for the transmission of a 1-bit, $\mathcal{B}(M,p_j)$ denotes the binomial distribution with $M$ trials and a success probability of $p_j$, and $\mathcal{N}(0,\sigma_n^2)$ is a Gaussian distribution with mean $0$ and variance $\sigma_n^2$ modelling the counting noise inside the receiver. Using the formula for the mean of a binomial distribution, $(b_i \cdot M \cdot p_i)$ gives the expected number of molecules released in the $(i-1)^{th}$ previous slot and detected in the current interval. For instance, $(b_1 \cdot M \cdot {p_{1}})$ is the expected number molecules that are both released and detected in the current signal slot. For sufficiently large values of the aggregate mean and variance of $N_{exp}$, the binomial sum at (3) can be approximated by a Gaussian normal distribution \cite{gaussian} using central limit theorem:

\begin{align*}
    N_{exp} \hspace{-1pt}&\sim\hspace{-1pt} \mathcal{N}\hspace{-1pt} \Big(\hspace{-1pt} \sum_{j=1}^{L} b_j \!\cdot\! \hspace{-1pt} M \!\cdot\! p_j, \hspace{-1pt}\big(\sum_{j=1}^{L} b_j \!\cdot\! M \!\cdot\! p_j \!\cdot\! (1 - p_j)\big)\hspace{-3pt} + \hspace{-1pt}\sigma_n^2 \hspace{-1pt}\Big)\hspace{-3pt}\hspace{8pt} \text{(4)}
\end{align*}

Since a sum of independent binomial random variables can be analytically non-trivial, the normal distribution given in (4) proves to be highly beneficial in many analytical derivations as we demonstrate in Section \ref{sec:estimation}. We will not use (4) for channel simulations.

\section{RLIM Coding and Codebook Derivation}
\label{sec:generalisedcoding}

 We propose a new infinite family of codebooks, denoted by RLIM$_{i}(n,k)$, where $i, n, k \in \mathbb{N}$. Here, RLIM stands for run-length-limited ISI-mitigation. We first define the properties of RLIM$_{i}(n)$ as follows.

\begin{enumerate}[label=\arabic*.] 
    \item Each 1-bit is followed by $i$ 0-bits, provided $i$ or more available positions exist following the 1-bit. If fewer than $i$ positions are available after the 1-bit, all subsequent bits are 0-bits.
    \item Each code starts with $i$ 0-bits.
    \item Each code must contain at least one 1-bit.
\end{enumerate}

 The 1$^{st}$ and 2$^{nd}$ restrictions are needed to ensure that, after an accurate detection of a single 1-bit of any code, none of the following $i$ bits can be a 1-bit. So that, if one of these $i$ bits are erroneously detected to be a 1-bit, it can be error corrected to a 0-bit. The 3$^{rd}$ condition has to do with the adaptive-threshold detection technique as used in \cite{best_channel_coding_2020}; and why it is needed will be illustrated in Section \ref{sec:error}. If the 3$^{rd}$ condition were dropped, RLIM$_{i}(n)$ would be equal to run-length-limited codes of order ($i,\infty$) of length $n$ with leading merging zeros \cite{RLL}.  Note that, when the order $i$ of the RLIM scheme is set to 1, resultant codes correspond to the ISI-mitigating codes \cite{best_channel_coding_2020}.

\label{sec:second}

 Let $C_{i}(n)$ denote the subset of all codes belonging to $\{0, 1\}^n$, with the 1$^{st}$ condition in place, but the 2$^{nd}$ and 3$^{rd}$ conditions dropped. In coding literature, $C_{i}(n)$ is known as a $d$-limited sequence (where $d$ substitutes $i$) \cite{d-limited}. Then, through basic Combinatorics, and from \cite{d-limited}, $C_{i}(n)$ conforms to the following relation (5), where each row represents a distinct code, and $|.|$ is the cardinality function. Note that $0^{m}_n$ denotes a matrix of $0$s with $m$ columns and $n$ rows; and similarly $1^{m}_n$ denotes a matrix of $1s$ with $m$ columns and $n$ rows.

\begin{equation*}
\begin{aligned}
\mathbf{C}_{i}(n) &\hspace{-2pt}= \hspace{-3pt}\left[
\begin{array}{c:c}

  \multicolumn{1}{c}{\hspace{-10pt}\boldsymbol{0_{\lvert C_{i}(n-1)\rvert}^{1}}}
  &\hspace{-20pt}\mathbf{C}_{i}(n-1)\\[1pt]
  \hdashline

  \multicolumn{1}{c}{\hspace{-4pt}\boldsymbol{%
      1_{\lvert C_{i}(n-1-i)\rvert}^{1}\,
      0_{\lvert C_{i}(n-1-i)\rvert}^{i}}}
  & \mathbf{C}_{i}(n-1-i)\hspace{-3pt}
\end{array}
\right] \\[-2pt]
& 
\end{aligned}
\tag{5}
\end{equation*}

\begin{tikzpicture}[overlay, remember picture]

  \draw[dashed, line width=0.5pt] (4.9, 1.8) -- (4.9, 2.5);
\end{tikzpicture}

\begin{tikzpicture}[overlay, remember picture]

  \draw[dashed, line width=0.5pt] (5.5, 1.7) -- (5.5, 2.3);
\end{tikzpicture}

\vspace{-0.8cm}

This recursive relation holds, because the elements of $C_{i}(n)$ can be categorized into two groups: those that start with a 0-bit, and those that start with a 1-bit. If a code starts with a 0-bit, the remaining part of it can be any element of $C_{i}(n-1)$. Similarly, If a code starts with a 1-bit, this 1-bit must be followed by $i$ 0-bits, then the remaining part can be any element of $C_{i}(n-1-i)$. To obtain the  RLIM$_{i}(n)$, first we need to enforce the 2$^{nd}$ condition, which was dropped in the derivation of $C_{i}(n)$. That is, $i$ 0-bits are concatenated to the start of each code of $C_{i}(n-i)$. Then, to enforce the 3$^{rd}$ condition, we need to remove the code that consist entirely of 0-bits from the matrix. These operations can be done via the matrix given in (6). Note that the superscript (*) indicates the removal of the code consisting completely of 0-bits.

\begin{equation}
\bold{RLIM}_{\boldsymbol{i}}\boldsymbol{(n)} = \left[
\begin{array}{c:c}

  \multicolumn{1}{c:}{{\mathbf{ 0_{\boldsymbol{|}C\boldsymbol{^{*}_{i}(n-i)|}}^{\boldsymbol{i}}}}} & \multicolumn{1}{c}{\mathbf{C}_{\boldsymbol{i}}^{\boldsymbol{*}}\boldsymbol{(n-i)}} \\
\end{array}
\right] \tag{6}
\end{equation}

\vspace{5pt}
As an example, we now derive the code space of RLIM$_{2}(6)$ using (5) and (6). We first obtain the following:

\begin{equation*}
C_{2}(1) =  \begin{bmatrix} 0 \\ 1 \end{bmatrix} \hspace{-0.09cm} , \hspace{0.11cm} 
C_{2}(2)  = \begin{bmatrix} 0 & 0 \\ 0 & 1 \\ 1 & 0 \end{bmatrix} \hspace{-0.09cm} , \hspace{0.11cm}
C_{2}(3)  = \begin{bmatrix} 0 & 0 & 0 \\ 0 & 0 & 1 \\ 0 & 1 & 0 \\ 1 & 0 & 0 \end{bmatrix} \tag{7}
\end{equation*}

Now apply the recursion matrix at (5) to obtain $C_{2}(4)$.

\begin{equation}
C_{2}(4) = \left[
\begin{array}{c:c}
  \multicolumn{1}{c}{\hspace{-16pt} 0_{4}^{1}} & \hspace{-16pt} C_{2}(3) \vspace{1pt}\\
  
  \hdashline
  1_{2}^{1}\ 0_{2}^{2} & C_{2}(1)
\end{array}
\right] = \begin{bmatrix} 0 & 0 & 0 & 0 \\ 0 & 0 & 0 & 1 \\ 0 & 0 & 1 & 0 \\ 0 & 1 & 0 & 0 \\ 1 & 0 & 0 & 0 \\ 1 & 0 & 0 & 1 \end{bmatrix} \tag{8}
\end{equation}

\begin{tikzpicture}[overlay, remember picture]

  \draw[dashed, line width=0.5pt] (2.75, 2.1) -- (2.75, 2.8);
\end{tikzpicture}

To get RLIM$_{2}(6)$, apply (6) to $C_{2}(4)$ as follows: 

\begin{equation}
\text{RLIM}_{2}(6) = \left[
\begin{array}{c:c}

  \multicolumn{1}{c:}{ 0^2_5 } & \multicolumn{1}{c}{C^{*}_{2}(4)} \\
\end{array}
\right] = \begin{bmatrix}  0 & 0 & 0 & 0 & 0 & 1 \\ 0 & 0 & 0 & 0 & 1 & 0 \\ 0 & 0 & 0 & 1 & 0 & 0 \\ 0 & 0 & 1 & 0 & 0 & 0 \\ 0 & 0 & 1 & 0 & 0 & 1 \end{bmatrix} \tag{9}
\end{equation}

Any RLIM$_{i}(n)$ can be obtained using the recursive procedure defined in this section. For a given value of $i$, only the base cases, $C_{i}(1), \ldots, C_{i}(j), \ldots, C_{i}(i+1)$, where $1 \leq j \leq i+1$, should be pre-computed as in (7). This is a trivial task: Each base-case $C_{i}(j)$ has, in total, $j+1$ codes (i.e., rows). The first code of each $C_{i}(j)$ is the $0$ vector, and the remaining $j$ codes have the following property: The $k^{th}$ code (i.e. row) of $C_{i}(j)$, where $2 \le k \le j+1$, has a 1-bit in the $(j+2-k)^{th}$ position, with remaining places all assigned with 0-bits.

Finally, to perform encoding and decoding, after deriving the code space RLIM$_{i}(n)$, a bijective function needs to be created between $\{0, 1\}^{\lfloor\log_2 |RLIM_{i}(n)|\rfloor}$ and a subset $S$ of RLIM$_{i}(n)$ with size $2^{\lfloor\log_2 |RLIM_{i}(n)|\rfloor}$, where $\lfloor.\rfloor$ is the floor function. The subset $S$ should include codes of RLIM$_{i}(n)$ with the fewest number of 1-bits. This criterion is important because it ensures that the coding scheme can utilize a greater number of information molecules for transmitting each 1-bit \cite{mec_codes}, thereby reducing the bit error rate. 

Let us denote such a subset S of RLIM$_{i}(n)$, by RLIM$_{i}(n,k)$ provided that $|RLIM_{i}(n)|\ge 2^k$. That is, $S$ a minimal subset with size $2^k$ of RLIM$_{i}(n)$ in term of the total number of 1-bits it contains. Since there can be multiple such minimal subsets, we choose the one that contains  lexicographically (in binary order) the smallest codes. After creating the subset $S$, we order it by the binary values of its codes to facilitate the binary search algorithm during decoding. Specifically, when converting each code in $S$ back to its associated element in $\{0,1\}^{k}$, the binary search enables retrieval in $\mathcal{O}(k)$ time. The Python implementations of codebook generation, encoding, and decoding algorithms, for any RLIM$_i(n,k)$, are given in Code Availability Section. For a RLIM$_i(n,k)$ codebook with $2^k$ codewords, the information rate is $R=k/n$, which (as $n$ grows) approaches the classical $(i,\infty)$ RLL capacity $C_i=\log_2\lambda_i$, where $\lambda_i>1$ is the Perron root of $x^{i+1}=x^i+1$~\cite{RLL}.

Classical $(d,k)$ RLL coding is well established \cite{RLL}.  Weight-aware variants also exist where the codebook is defined by fixing the Hamming weight or bounding the running digital sum, and then indexing those sets enumeratively \cite{the_three,the_four}. Since the effect of the $3^{rd}$ restriction is minuscule, what actually sets apart RLIM codes from conventional RLL codes is the minimisation of the total number of 1-bits in the code space.  Our proposed RLIM forms a fixed-rate, minimum-weight subset drawn across multiple weight classes, rather than taking a single constant-weight (or bounded-weight) class. To the best of our knowledge, this fixed-size minimum-weight selection principle for $(i,\infty)$ constraints has not been explicitly formulated in prior work \cite{RLL,the_three,the_five}. As we will demonstrate in the performance evaluation section, this selection will prove advantageous over the classical RLL codes, whose \(2^k\) codewords are typically chosen lexicographically (as in our simulation comparisons), without any 1-bit minimisation.

\subsection{Detection}

\label{sec:error}

Assume that an RLIM$_{i}(n)$  coding scheme is used. We will adopt the adaptive threshold detection technique. Let \(m=(m_1,\ldots,m_n)\) be the sequence giving the absorbed number of molecules at each signal interval of a code $c\in$ RLIM$_{i}(n) $. Define $m_{max}=$$max\{m_{i+1},...,m_n\}$\footnote{For ISI-mitigating codes, $m_{max}$ is defined as the maximum of $m - \{m_1\}$ (i.e., the sequence $m$ excluding its first element). By redefining $m_{max}=max\{m_{i+1},...,m_n\}$, we generalize this approach beyond ISI-mitigating codes.} and $m_{min}=$$min\{m_{i+1},...,m_n\}$\footnote{For ISI-mitigating codes, $m_{min}$ is defined as the minimum of $m - \{m_1\}$. By redefining $m_{min}=$$min\{m_{i+1},...,m_n\}$, we generalize this approach beyond ISI-mitigating codes.}. Then the adaptive threshold, $\tau^{m}_{adapt}$, pertaining to $m$ is found as  
\begin{equation*} \tau^{m}_{adapt} = a\cdot m_{min} + (1-a)\cdot m_{max}, \hspace{0.2cm} \tag{10}\vspace{-2pt}\end{equation*} where $a$ is the channel-specific scaling constant \cite{best_channel_coding_2020}. Note that $0\leq a \leq1$, and thus $m_{min}\leq \tau^{m}_{adapt} \leq m_{max}$. If the number of absorbed molecules falls below $\tau^{m}_{adapt}$, the corresponding signal interval is detected as a 0-bit, otherwise it is detected as a 1-bit. To find the optimal value of $a$, ensuing pilot signals, whose content is pre-known by the receiver, are sent \cite{best_channel_coding_2020}; and the value of $a$  that results in the least BER value is chosen. The receiver can do this selection by increasing $a$ from $0$ to $1$ with a pre-determined step size of $s$, whose value is taken to be $0.005$. Through this approach, an element of $m$ that has the value $m_{max}$ is always detected to be a 1-bit. This is why RLIM$_{i}(n)$ needs at least one 1-bit in each code.

\begin{algorithm}[t]
  \caption{Proposed Detection with Static Threshold}
  \begin{algorithmic}[1]
    \Require Let $m$ be the sequence of absorbed‐molecule counts of length $n$,
             $\tau_{\mathrm{static}}$ the static threshold, and $i$ the order of
             $\text{RLIM}_i(n,k)$.
    \State Let $detected\_bit\_sequence \gets \mathbf{0}_{1\times n}$

    \For{$j \gets 1$ \textbf{to} $n$}
        \If{$m[j] \ge \tau_{\mathrm{static}}$}
            \State $detected\_bit\_sequence[j] \gets 1$
        \EndIf
    \EndFor

    \If{$detected\_bit\_sequence[i+1:n]$ is the $0$-vector}
        \State $l \gets \displaystyle\arg\max_{h = i+1,\dots,n} m[h]$
        \State $detected\_bit\_sequence[l] \gets 1$
    \EndIf

    \State \textbf{return} $detected\_bit\_sequence$
  \end{algorithmic}
\end{algorithm}

\begin{algorithm}[t]

\caption{Proposed Error Correction for RLIM}

\begin{algorithmic}[1]  
\Require  $detected\_code$ with size $n$,  order $i$ of RLIM$_i(n,k)$

\State{ $j = 1$; $skip = i$ }
\While{$j \le n$}
    \If{$skip > 0$}                     
        \State $detected\_code[j] = 0$
        \State $skip = skip - 1$
    \Else                                
        \If{$detected\_code[j] == 1$}
            \State $skip = i$       
        \EndIf
    \EndIf
    \State $j = j + 1$
\EndWhile

\end{algorithmic}
\end{algorithm}

Additionally, we provide the implementation of RLIM-specific static-threshold detection technique in Algorithm 1, as we will evaluate both the static and dynamic approaches in Section \ref{sec:performance}. Algorithm 1 strengthens the classical static threshold algorithm \cite{BCSK} by providing robustness against edge cases through its Lines 7–10 \footnote{This holds when the detected molecule-count sequence contains a non-zero value. In our implementation, whenever the detected integer sequence is the 0 vector, we return a code whose only 1-bit is located in its $(i+1)$th index (where the indices start from 1).}. It ensures that each detected code contains at least one 1-bit, which holds true for all RLIM$_{i}(n)$ codes.  Using pilot signals, the optimal static threshold for a given channel can be determined similarly to the optimal scaling constant derivation: By testing thresholds from 1 to $M$, the threshold that yields the lowest BER is identified.

\subsection{Error Correction}
\label{sec:error_correct}

After detection, a resultant binary sequence may not be an element of the coding scheme used. Thus, error correction is needed. By observing that in any code of $\mathrm{RLIM}_{i}(n)$, the first $i$ bits are $0$ and no $1$-bit can be followed by another $1$-bit within the next $i$ positions, we propose Algorithm 2 for error correction. Note that, when the order $i$ is set to $1$ in Algorithm 2 the resulting algorithm coincides with the one for ISI-mitigating codes presented in \cite{best_channel_coding_2020}. To the best of our knowledge, we are not aware of any prior \(O(n)\) time decoder that, for any $y\in\{0,1\}^n$, returns a minimum-Hamming-distance sequence satisfying the general $(i,\infty)$–RLL constraint (with leading $i$ zeros); Algorithm 2 provides such a procedure. After applying the error correction algorithm to a detected code, the resultant code $c$ may not be an element of RLIM$_{i}(n,k)$, in which case we iteratively substitute the right-most 1-bit of the code $c$ with a 0-bit, while searching it inside RLIM$_{i}(n,k)$ at each substitution. We now state the following theorem, which establishes the optimality of our RLIM error‐correction algorithm.

\textit{Theorem:} In terms of Hamming distance (the number of differing bits between two binary words of equal length), for \((i,\infty)\)-RLL codes, Algorithm~2 is optimal. That is,

\begin{equation}\tag{11}
\begin{aligned}
\forall\,y\in\{0,1\}^n:\;
d_H\!\bigl(F_i(y),y\bigr)
&=\\ \min\Bigl\{\, d_H(x,y):
& \hspace{4pt} x\in RLL_i(n) \Bigr\}.
\end{aligned}
\end{equation} 
where \(F_i(y)\) denotes the output of Algorithm~2, \(d_H\) is the Hamming distance, and $RLL_i(n)$ is the set of all RLL codes of length $n$ and order $i$ with leading 0-bits.

\textit{Proof:} Let \(y\in\{0,1\}^n\).
If \(x\in RLL_i(n)\) and for some \(t\) we have \(y_t=0\) and \(x_t=1\), define \(x'\) by setting \(x'_t=0\) and \(x'_u=x_u\) for \(u\neq t\).
Then \(x'\in RLL_i(n)\) (replacing a \(1\) by \(0\) cannot violate the RLL constraint) and \(d_H(x',y)=d_H(x,y)-1\). Therefore if $x^*\in RLL_i(n) $ has the smallest $d_H(x^*,y)$ (i.e, $x^*$ is a minimizer code), then the indices of 1-bits of $x^*$ (denoted as $ones(x^*)$) are a subset of $\{\, t \in \{i+1,\dots,n\} : y_t = 1 \,\}$. We map each $1$-bit of $y$ at index $t\ge i{+}1$ to the half-open interval $[t,\,t+i+1)$. We are then trying to choose the maximum number of $1$-bits with non-overlapping intervals to minimize the Hamming distance. Selecting $1$-positions that satisfy the \((i,\infty)\) constraint is exactly selecting a maximum-size set of non-overlapping intervals, for which the earliest-feasible greedy rule is optimal (\cite{CLRS}, Chapter~16.1).
Algorithm~2 implements this rule by scanning left-to-right, keeping the earliest feasible $1$ and skipping the next $i$ positions; hence it maximizes $|ones(x)|$, thus minimizing $d_H(x,y)$. \qed

Note that, $RLIM_i(n) \subset RLL_i(n)$. Since our detection methods guarantee the existence of at least one \(1\)-bit in positions \(i+1\) through \(n\) of a detected binary sequence \(c\) of length $n$,  $F_i(c)$ is a RLIM$_i(n)$ code by the construction of the Algorithm 2. Therefore, the theorem above also applies for the RLIM codebooks, resulting in the following corollary:
\textit{Corollary:} In terms of Hamming distance, for $RLIM_i(n)$ codes, Algorithm~2 is optimal. That is,

\begin{equation}\tag{12}
\begin{aligned}
\forall\,y\in\{0,1\}^n,\;
\bigl[\exists\, t \in \{i{+}1,\dots,n\}: y_t = 1\bigr] 
\Longrightarrow\;
\hspace{-4cm} \\  d_H\!\bigl(F_i(y),y\bigr)
&= \min_{x\in RLIM_i(n)} d_H(x,y).
\end{aligned}
\end{equation}

\vspace{0.25cm}

\subsection{Equivalence with Viterbi Decoding and Complexity}

We now present the generic, standard trellis-based error-correction procedure for $(i,\infty)$–RLL codes, known as Viterbi hard-decoding \cite{mitViterbiNotes, original_viterbi}, which selects the codeword of minimum Hamming distance to the detected word. Let $y\in\{0,1\}^n$ be the detected word and let $x\in\{0,1\}^n$ be a candidate codeword. We decode on the standard $(i,\infty)$–RLL constraint as a labeled digraph with vertices $\{0,\dots,i\}$ \cite{RLL}: from each state (i.e., vertex) $s$, there is exactly one outgoing edge labeled $0$-bit to $\min\{s+1,i\}$, and only from state $i$ there is an outgoing edge labeled $1$-bit to the state $0$. A length-$n$ path from start state $s_0=0$ (implementing the leading merging 0-bits, i.e., the first $i$ outputs are $0$-bits) yields $x$ by reading edge labels; termination can happen at any state. 

With Hamming branch cost $\mathbf{1}[x_t\neq y_t]$, the add–compare–select (Viterbi) recursion returns a minimum-Hamming-distance codeword $x$ to $y$. Standard Viterbi expositions leave equal-metric ties unspecified (‘break ties arbitrarily’) \cite{mitViterbiNotes}, and real implementations resolve equality with a fixed comparator policy. In hardware/software, this typically amounts to using a strict  comparator (always pick the first candidate when equal), a non-strict comparator (pick the later candidate when equal), or a randomized tie-break (select uniformly among equal-cost candidates at each state/time and among equal-cost terminal states) \cite{two_viterbi}.

With the deterministic Viterbi last-wins policy (i.e., in case of ties at state \(i\), prefer the self-loop \(i\xrightarrow{\,0\,} i\) over \((i\!-\!1)\xrightarrow{\,0\,} i\), and at termination choose the tied state with the largest index), RLIM Algorithm~2 and Viterbi yield identical outputs; this follows by a short induction on the trellis time index over the Viterbi recursion. RLIM runs in \(O(n)\) time with \(O(1)\) auxiliary space. Viterbi runs in \(O((i{+}1)n)\) time with \(O((i{+}1)n)\) space using full back-pointers (or \(O(i{+}1)\) auxiliary space with checkpointed or divide-and-conquer traceback). Consequently, Algorithm 2 is a better choice than its Viterbi-equivalent.

\begin{figure}[t]
  \centering

  \includegraphics[width=1\columnwidth]{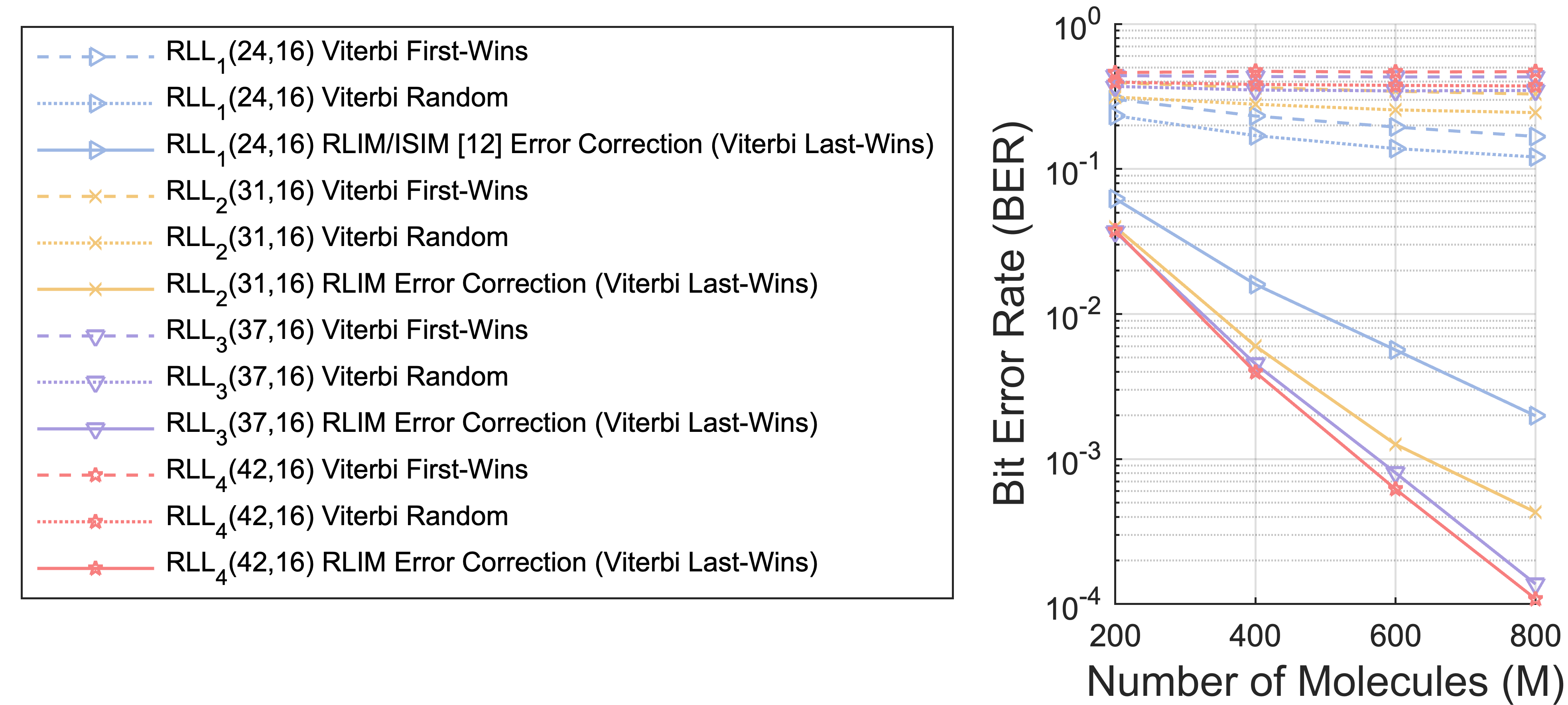}
  \caption{Error Correction Comparisons for RLL Codes with $t_s=200$ $ms$, $\sigma_n^2=0$, and $r_0=$	 $10\,\, \si{\micro\metre} $ }
  \label{fig:onecol}
\end{figure}

We therefore compare three decoders for \((i,\infty)\)-RLL: generic Viterbi (first-wins, random tie-breaking) and our RLIM error correction (equivalent to Viterbi last-wins). Simulation parameters, other than those listed in the caption of Fig. 2, are listed in Tables I–III, and Section IV details the simulation rationale and procedure. We use static-threshold detection, which yields the best BER under the non-drift channel settings of Section IV. For each comparison, we first transmit $46080$ bits by encoding six independent $7680$-bit sequences to determine the optimal threshold. Using those thresholds, we then send $967680$ bits (six contiguous runs of $161280$-bit encodings) to estimate BER. As shown in Fig.~2, Viterbi with generic tie-breaking (first-wins or random) yields higher BER, whereas RLIM (i.e., Viterbi last-wins) achieves the lowest BER. Accordingly, for the remainder of this work we adopt the RLIM error-correction algorithm when benchmarking classical RLL and for RLIM codes themselves. There is a thesis \cite{RLL_thesis}, where RLL codes(including order ($1, \infty$)) have been applied to MC. However, the author obtained numerical results in which RLL codes were surpassed by uncoded transmissions in terms of BER. With Algorithm 2 implemented to RLL codes, in the performance evaluation section, we will show that RLL codes gain a significant advantage over uncoded transmissions.

\subsection{Analytical Estimation of Static Threshold}
\label{sec:estimation}

Both the adaptive and static threshold techniques described in the previous section require sending pilot signals at the start of the communication. However, if the receiver has the knowledge of the channel parameters, transmitting pilot signals beforehand may no longer be necessary. 

Our contribution in this subsection is adapting the analytical derivation method for ISI-mitigating codes \cite{best_channel_coding_2020} to RLIM$_{i}(n)$, with significant modifications which will be clarified at the end of this subsection. Define $ ISI_j = \mathcal{N}(M \cdot p_j, M \cdot p_j \cdot (1 - p_j)) $ as per (4), which approximates the distribution of the expected number of molecules detected from the emission of $ M $ molecules at the $ (j-1)^{th} $ previous signal slot. Assuming RLIM$_i(n) $ is used, when a 1-bit is transmitted in the current signal slot, the distribution of the maximum expected number of detected molecules can be given by the (13), where $\mathcal{N}(0,\sigma_n^2)$ denotes the Gaussian noise.

\[
 {}^iN_1^{max}= \lim_{I_1 \to \infty} \Big(\sum_{k=1}^{I_1} ISI_{1 + (i + 1) \cdot (k - 1)}\Big)+\mathcal{N}(0,\sigma_n^2) \tag{13}
\]

\vspace{0.1cm}

\begin{figure}[t]    
    \centering
    \begin{subfigure}[b]{0.97\linewidth}
        \centering
        \includegraphics[width=\linewidth]{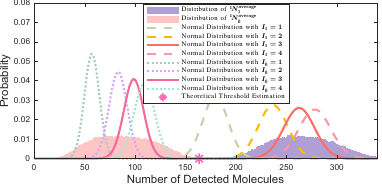}
        \caption{RLIM$_{1}(24,16)$ \cite{best_channel_coding_2020} with $M=1294$ and $t_s=((16/24)\cdot 200)$ ms}
        \label{fig:subfig1}
    \end{subfigure}

    \vspace{1em}
    
    \centering
    
    \begin{subfigure}[b]{0.97\linewidth}
        \centering
        \includegraphics[width=\linewidth]{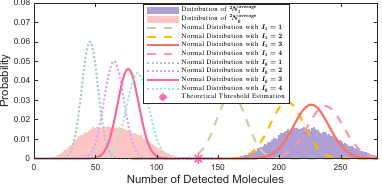}
        \caption{RLIM$_{2}(31,16)$  with $M=1484$ and $t_s=((16/31)\cdot 200)$ ms}
        \label{fig:subfig2}
    \end{subfigure}

\vspace{1em}
    
    \centering

    \begin{subfigure}[b]{0.97\linewidth}
        \centering
        \includegraphics[width=\linewidth]{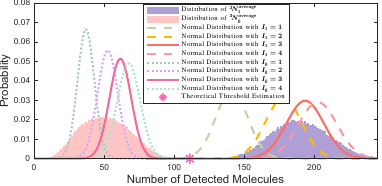}
        \caption{RLIM$_{3}(37,16)$ with $M=1590$ and $t_s=((16/37)\cdot 200)$ ms}
        \label{fig:subfig3}
    \end{subfigure}

\vspace{1em}
    
    \centering

\begin{subfigure}[b]{0.97\linewidth}
        \centering
        \includegraphics[width=\linewidth]{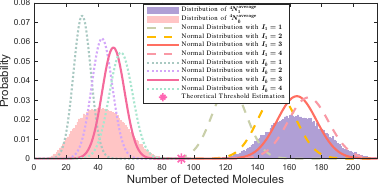}
        \caption{RLIM$_{4}(42,16)$ with $M=1621$ and $t_s=((16/42)\cdot 200)$ ms}
        \label{fig:subfig4}
    \end{subfigure}

    \caption{Detected Molecule Distributions for coding schemes RLIM$_i(n_i,16)$ with, $ D = 79.4 $ $\si{\micro\metre^2/\second} $, $ r_R = 5$ $ \si{\micro\metre} $, $ r_0 = 10$ $ \si{\micro\metre} $, $L$ $= 200$, $\sigma_n^2=0$, and unnormalized signal interval and molecule counts of $ t_s = 200 $ ms, and  $ M = 1000 $. This figures motivates $I_1=I_{\hat{0}}=3$ in (14) and (16).}
 
\end{figure}

Rather than deriving a distribution for the maximum possible number of detected molecules, we aim to derive a distribution for the average case. To approximate mean value of a such a distribution, $ I_1 $ should be a positive integer. To illustrate this phenomenon, consider the detection count distribution from a binomial simulation of $10000$ consecutive codewords, each encoding a random bit sequence of length 16. The simulation uses RLIM codes of orders from 1 to 4 in an MC channel whose simulation parameters are provided in the caption of Fig. 3, with corresponding normalized parameters given in the sub-captions. The simulator selection and normalization procedure will be explained in the next subsection. As can be seen in the right sections of each subfigure in Fig. 3, $I_1 = 3$ provides a more accurate representation of the molecule count distribution of 1-bits (shown in purple) compared to other values of $I_1$, based on its similarity to the mean value of the distribution. Please note that we observed a similar phenomena for molecule count ($M$) values smaller than $1000$. Accordingly, we assume the following approximation in (14):

\[
{}^iN_{1}^{average} \approx \Big(\sum_{k=1}^{3} ISI_{1 + (i + 1) \cdot (k - 1)}\Big) +\mathcal{N}(0,\sigma_n^2) \tag{14}
\]

Now, consider the case where the current signal slot corresponds to a 0-bit, and all preceding signal slots have been detected accurately. For any code of RLIM$_i(n)$, this implies that if there are any 1-bits within $ i $ previous signal slots or if the current interval corresponds to one of the first $ i $ positions of a code, then regardless of the threshold value, the error correction algorithm will always correctly detect the current signal slot as a 0-bit. Accordingly, we denote a 0-bit, which is neither in one of the first $ i $ positions of a code nor has any preceding 1-bits within $ i $ slots, as a $ \hat{0} $-bit. For the transmission of a $ \hat{0} $-bit, the maximum expected number of detected molecules is given by the following distribution:

\[
{}^iN_{\hat{0}}^{max} =\lim_{I_{\hat{0}} \to \infty} \Big(\sum_{k=1}^{I_{\hat{0}}} ISI_{1+(i + 1) \cdot (k)}\Big)+\mathcal{N}(0,\sigma_n^2) \tag{15}
\]

Setting $I_{\hat{0}} = 2$ provides a better approximation for the mean of ${}^iN_{\hat{0}}^{\text{average}}$ (as shown in the left sections of the subfigures in Fig.~3), but, like other values of $I_{\hat{0}}$, it fails to accurately model the variance. Moreover, $I_{\hat{0}} = 2$ is insufficient for predicting the distribution of ${}^iN_{\hat{0}}^{\text{average}}$ at higher molecule counts. We chose $I_{\hat{0}} = 3$, as given in (16), as it more accurately approximates the distribution at larger detected molecule counts.

\[
{}^iN_{\hat{0}}^{average} \approx \Big(\sum_{k=1}^{3}{ ISI_{1+ (i + 1) \cdot (k)}\Big)}+ \mathcal{N}(0,\sigma_n^2)\tag{16} 
\]

Future research should focus on developing more accurate theoretical distributions for ${}^iN_1^{average}$ and ${}^iN_{\hat{0}}^{average}$. While the complex and recursive nature of RLIM$_i(n,k)$ makes theoretically obtaining these distributions a challenging task, achieving more precise distributions will lead to BER-wise improved static threshold values. This, when the channel parameters are known, will further discard the need to send pilot signals in advance of the communication.

Recall that we defined $ ISI_j = \mathcal{N}(M \cdot p_j, M \cdot p_j \cdot (1 - p_j)) $ as per (4), where \(M\) and \(p_j\) are as defined in (2)–(4). For simplicity, we make the following substitutions:

\begin{align}
A &= \left( \scalebox{0.89}{$\sum_{k=1}^{3}$} M \cdot (p_{1+(i + 1) \cdot k}) \right) \nonumber \\
B &= \left( \scalebox{0.89}{$\sum_{k=1}^{3}$} M \cdot (p_{1+(i + 1) \cdot k}) \cdot (1 - p_{1+(i + 1) \cdot k}) \right) + \sigma_n^2 \nonumber \\
C &= \left( \scalebox{0.89}{$\sum_{k=1}^{3}$} M \cdot p_{1 + (i + 1) \cdot (k - 1)} \right) \nonumber \\
D &= \left( \scalebox{0.89}{$\sum_{k=1}^{3}$} M \cdot (p_{1 + (i + 1) \cdot (k - 1)}) \cdot (1 - p_{1 + (i + 1) \cdot (k - 1)}) \right) + \sigma_n^2 \vspace{0.0cm}\tag{17} 
\end{align}

In (17), A and B are respectively the mean and the variance of (16) and C and D are respectively the mean and the variance of (14). Therefore, the distribution (14) can be denoted as $\mathcal{N}(C,D)$ and the distribution (16) can be denoted as $\mathcal{N}(A,B)$. In literature, $Q$ function is defined as the tail probability of a Gaussian normal distribution (i.e., $Q\big(\tfrac{x - a}{\sqrt{b}}\big)$ gives the integration of $\mathcal{N}(a,b)$ from $x$ to $\infty$). Let us denote $\tau_{static}$ as the chosen static threshold constant. If we assume that the current transmitted bit is a 1-bit, correctly detecting the current signal interval has a probability of $Q\big(\tfrac{\tau_{static} - C}{\sqrt{D}}\big)$. If the current transmitted bit is a $\hat{0}$-bit, correctly detecting the current signal interval has a probability of $1-Q\big(\tfrac{\tau_{static} - A}{\sqrt{B}}\big)$. Let $P_1$ denote the appearance probability of 1-bits in the code space RLIM$_i(n,k)$, and similarly let $P_{\hat{0}}$ denote the appearance probability of $\hat{0}$-bits inside the code space RLIM$_i(n,k)$. Then, the probability of correctly detecting the current signal slot, assuming all preceding slots have been detected accurately, can be given as follows:

\begin{equation}
P \;=\; P_{\hat{0}} \cdot \left[ 1 - Q\!\left( \frac{\tau_{static} - A}{\sqrt{B}} \right) \right]
      \;+\; P_{1} \cdot \, Q\!\left( \frac{\tau_{static} - C}{\sqrt{D}} \right)
\tag{18}
\end{equation}

We need to find the highest value of (18) in terms of $\tau_{static}$. Then, taking the derivative of (18) with respect to $\tau_{static}$, and equating it to zero, the analytical derivation of $\tau_{static}$ is obtained as in (19). An algebraic derivation is provided in the Appendix.

\begin{align*}
\tau_{static} &= 
\scalebox{1.0}{$
\frac{D \cdot A - B \cdot C + \sqrt{B \cdot  D \cdot   \Big(\hspace{-1pt} \hspace{-1pt} (C - A)^2 - 2 \cdot (B - D) \cdot \log_{e} \hspace{-1pt} \Big( \frac{\sqrt{D} \cdot P_{\hat{0}}}{\sqrt{B} \cdot P_1} \Big)\Big)}}{D-B}\hspace{5pt}
$}
\begin{array}{c}
\vspace{-8ex} \\ 
\end{array}  \tag{19}
\\
& \hspace{210pt}
\end{align*} As an example, for RLIM$_4($42$,16)$, the values of P$_{\hat{0}}$ and P$_1$ are $\frac{996497}{42\cdot2^{16}}$ and $\frac{323397}{42\cdot2^{16}}$, respectively. Then, we calculate the corresponding $\{p_x\}$ values from (2) based on the normalized channel parameters outlined in the sub-caption of Fig. 3(d) and the remaining parameters that are given in the main caption of Fig. 3. Substituting all these values into (19) yields a static threshold estimate of $\approx$$92.13$, as marked in Fig. 3(d).

In \cite{best_channel_coding_2020}, the dynamic threshold formula $( a \cdot m_{min} + (1-a) \cdot m_{max} )$ is used in place of $\tau_{static}$ within (19). The values of $I_1$ and $I_{\hat{0}}$ are both set to 1. The derivative of (19) is then taken with respect to $a$, set equal to zero, and solved for $a$. As a result, in \cite{best_channel_coding_2020}, a new analytical value for $a$ is determined for each detected codeword molecule count sequence, $m$, based on $m_{max}$ and $m_{min}$. The dynamic threshold, $\tau^{m}_{dynamic}$, is updated accordingly, as per (10).

However, in such an analytical approach, regardless of the values $m_{max}$ and $m_{min}$ take, $\tau_{dynamic}^{m}$ always equals $\tau_{static}$. Therefore, while our approach in this paper for RLIM$_1(n)$ is equivalent to that in \cite{best_channel_coding_2020} for ISI-mitigating codes  when $I_{\hat{0}}=1$ and $I_1=1$, it is computationally more efficient: Our method requires computation only once and can be applied across all codeword detections. Moreover, as demonstrated in Fig. 3, the values $I_1=3$ and $I_{\hat{0}}=3$ generally provide a better estimate than $I_1=1$ and $I_{\hat{0}}=1$. This is because, if $I_1=1$ and $I_{\hat{0}}=1$ were used, the threshold value would decrease in all four subfigures of Fig. 3; it is evident that this would result in a poorer estimate, as the threshold would intersect with the distribution of ${\hat{0}}$-bits, leading to a greater number of detection errors.

\section{Performance Evaluation}
\label{sec:performance}

There are 3 main approaches to simulate an MC channel. The first is a discrete particle-tracking-based simulator.  This method operates in discrete time steps (\(\Delta t\)), updating the 3D positions of all information molecules by independently drawing each coordinate from the normal distribution \begin{equation}
\mathcal{N}(0,2D\Delta t),
\tag{20}
\end{equation} derived from diffusion physics \cite{MC_diff,diffusion_normal_distribution}.  It then counts how many molecules are absorbed by the receiver during each time step and removes the absorbed molecules from the simulation. The second approach is to use the summation of binomial distributions expressed in (3) to compute the absorbed (i.e., detected) molecule count within a signal interval. The third and computationally fastest approach utilizes the normal distribution formula given in (4). Although this method is efficient, it may introduce slight errors for lower molecule counts and can occasionally produce negative detection counts. For simulating non-drift MC channel conditions, we employ the binomial distribution approach for its accuracy and computational efficiency with a channel memory of $200$\footnote{Using $F(t)=\frac{r_R}{r_0}\,\mathrm{erfc}\!\big((r_0-r_R)/\sqrt{4Dt}\big)$ with our parameter ranges ($D{=}79.4\,\si{\micro\metre^2/\second}$, $r_R{=}5\,\si{\micro\metre}$, $r_0\!\in\![9.5,11.5]\,\si{\micro\metre}$, and all (normalized) $t_s$ we use, including $0.2$ and $0.25$\,s), the largest $p_{200}=F(200t_s)-F(199t_s)$ occurs at $r_0{=}11.5\,\si{\micro\metre}$ and $t_s{=}(16/42)\cdot 0.2\,\mathrm{s}\approx 0.0762\,\mathrm{s}$, yielding $p_{200}\approx 1.14\times 10^{-4}$. Thus, even in this edge case, post-200 taps are negligible, justifying our use of 200 as the channel memory ($L$).}.

\subsection{Normalizations and Simulation Parameters}

To fairly compare different coding strategies, it is essential to normalize the signal interval and the molecule count per transmission of a 1-bit values. This ensures that the same amount of information is transmitted across different coding schemes within equal time periods and using an identical number of information molecules.The normalization, as given in \cite{moac} and \cite{normalization}, is done as follows: Let $I$ be the set of all available information. Assume each element of $I$ is equally likely to be transmitted. Suppose that a coding scheme $C_1$ encodes all the information of $I$, using $S_1$ bits, and $M_1$ 1-bits. Also assume that a coding scheme $C_2$ encodes all the information of $I$, using $S_2$ bits, and $M_2$ 1-bits. The signal interval value for coding scheme $C_2$ should be $S_1/S_2$ times that of the coding scheme $C_1$. Likewise, the number of molecules transmitted per 1-bit in coding scheme $C_2$ should be $M_1/M_2$ times that in coding scheme $C_1$. For our simulation, let $I$ be the set $\{0, 1\}^{16}$. To encode $I$ using RLIM$_i(n)$ we derive the following inequalities:

\begin{equation*}
    \hspace{10pt}\left|\text{RLIM}_{1}(23)\right| < 2^{16} < \left|\text{RLIM}_{1}(24)\right| = 75024 < 2^{17}
\tag{21}
\end{equation*}

\begin{equation*} 
\hspace{10pt}\left|\text{RLIM}_{2}(30)\right|<2^{16}<|\text{RLIM}_{2}(31)|=85625<2^{17}
\tag{22}\end{equation*} 

\begin{equation*} 
\hspace{10pt}|\text{RLIM}_{3}(36)|<2^{16}<|\text{RLIM}_{3}(37)|=82628<2^{17} \hspace{-0pt}
\tag{23}\end{equation*} 

\begin{equation*} 
\hspace{10pt}|\text{RLIM}_{4}(41)|<2^{16}<|\text{RLIM}_{4}(42)|=67984<2^{17} 
\tag{24}\end{equation*}

\begin{table}[!t]
\centering
\renewcommand{\arraystretch}{1.1}
	\caption{Normalized Signal Interval Values}

    \renewcommand{\arraystretch}{1}  
    \setlength{\tabcolsep}{5pt}  
    \label{tbl_system_parameters}
    \begin{tabular}{lcc}
        \hline
        \bfseries{Coding Method} & \multicolumn{2}{c}{\bfseries{Signal Interval of}} \\ 
        \cline{2-3}
         & \bfseries{\hspace{0pt}200 ms} & \bfseries{\hspace{16pt}250 ms} \\ 
        \hline
        Uncoded [16→16] & $200$ ms &\hspace{15pt} $250$ ms \\
        RLIM$_{1}(24,16)$ [16→24] \cite{best_channel_coding_2020} & $(16/24)\cdot 200$ ms &\hspace{15pt} $(16/24)\cdot 250$ ms \\
        RLIM$_{2}(31,16)$ [16→31] & $(16/31)\cdot 200$ ms &\hspace{15pt} $(16/31)\cdot 250$ ms \\
        RLIM$_{3}(37,16)$ [16→37] & $(16/37)\cdot 200$ ms &\hspace{15pt} $(16/37)\cdot 250$ ms \\
        RLIM$_{4}(42,16)$ [16→42] & $(16/42)\cdot 200$ ms &\hspace{15pt} $(16/42)\cdot 250$ ms \\
        Hamming(4,7)\hspace{1pt} [16→28] \cite{hamming} & $(16/28)\cdot 200$ ms &\hspace{15pt} $(16/28)\cdot 250$ ms \\
    ISI-Free(4,2,1) [16→32] \cite{ISI-Free} & $(16/32)\cdot 200$ ms &\hspace{15pt} $(16/32)\cdot 250$ ms \\
        \hline
        \vspace{0.1cm}
    \end{tabular}

	\caption{\hspace{-1pt}Number\hspace{-1pt}  of 1\hspace{-1pt}-bits and Normalized Molecule Counts } 

    \renewcommand{\arraystretch}{1.0} 
    \setlength{\tabcolsep}{4pt} 
    \label{tbl_combined_parameters}
    \begin{tabular}{l c c}
    \hline
    \bfseries Coding Method & \bfseries  Number & \bfseries Molecule Count \\ 
    & \bfseries of 1-bits & \bfseries ($M \in \mathbb{N}^+$) \\ 
    \hline 
    Uncoded [16→16] & $16 \cdot 2^{\scalebox{0.6}{$\boldsymbol{15}$}}=$ 524288 & $1 \cdot M$ \\
    RLIM$_{1}(24,16)$ [16→24] \cite{best_channel_coding_2020} & 405251 & $\left\lfloor 1.2937 \cdot M \right\rceil$ \\
    RLIM$_{2}(31,16)$ [16→31] & 353228 & $\left\lfloor 1.4842 \cdot M \right\rceil$ \\
    RLIM$_{3}(37,16)$ [16→37] & 329724 & $\left\lfloor 1.5900 \cdot M \right\rceil$ \\
    RLIM$_{4}(42,16)$ [16→42] & 323397 & $\left\lfloor 1.6211 \cdot M \right\rceil$ \\ 
    RLL$_{1}(24,16)$ [16→24] & 416350 & $\left\lfloor 1.2592 \cdot M \right\rceil$ \\
    RLL$_{2}(31,16)$ [16→31] & 370310 & $\left\lfloor 1.4158 \cdot M \right\rceil$ \\
    RLL$_{3}(37,16)$ [16→37] & 343276 & $\left\lfloor 1.5273 \cdot M \right\rceil$ \\
    RLL$_{4}(42,16)$ [16→42] & 325735 & $\left\lfloor 1.6095 \cdot M \right\rceil$ \\
    Hamming(4,7) \hspace{1pt}[16→28] \cite{hamming} & 917504 & $\left\lfloor 0.5714 \cdot M \right\rceil$ \\
    ISI-Free(4,2,1) [16→32] \cite{ISI-Free} & 1048576 & $\left\lfloor 0.5 \cdot M \right\rceil$ \\
    \hline
    \end{tabular}
    \vspace{4pt} 
	\renewcommand{\arraystretch}{1}
\end{table}
\begin{table}[!t]

	\begin{center}
	\caption{Simulation Parameters}

	\renewcommand{\arraystretch}{1.0} 
	\setlength{\tabcolsep}{4pt} 
	\label{tbl_system_parameters}
	\begin{tabular}{p{5.5cm} l}
	\hline
	\bfseries{Parameter} 							& \bfseries{Value}   \vspace{0.05cm} \\ 
	\hline
	Diffusion coefficient ($D$) 	& $79.4\,\,\si{\micro\metre^2/\second}$\\
	Distance between T$_x$ and R$_x$ ($r_0$)			& $9.5-11.5\,\, \si{\micro\metre} $\\
	Receiver radius ($r_R$)			& $5\,\,\si{\micro\metre}$\\
    Molecule count per a 1-bit for Uncoded ($M$)           & $100-1000$ \\
   Signal interval for Uncoded ($t_s$)           & $200-250$ ms \\
    Receiver Gaussian counting noise variance ($\sigma_n^2$)           & $0-20$ \\
    Channel Memory ($L$)  & $200$  \\
	\hline
	\end{tabular} 
	\end{center}
	\renewcommand{\arraystretch}{1}

\end{table}

\vspace{0.1cm}

Thus, each of the sets RLIM$_{1}(24,16)$, RLIM$_{2}(31,16)$, RLIM$_{3}(37,16),$ and RLIM$_{4}$ ${(42,16)}$ is a valid codebook for encoding the information set $\{0,1\}^{16}$. As explained previously, these codebooks are created in a way that minimizes the total number of 1-bits contained inside them. The normalized signal interval periods are given in Table I, for two different values: $200$ ms and $250$ ms. Table II presents the number of 1-bits in RLIM codebooks, along with the molecule count per a transmission of 1-bit values, normalized according to the Uncoded case, where $\left\lfloor . \right\rceil$ rounds the given number to the nearest integer. For instance, under our normalization, ISI-free $(4,2,1)$ halves $t_s$ and the per 1-bit molecule budget, increasing BER significantly. The full range of the parameters used in our comparative MC simulations are listed in Table III. The values of the parameters $D$, $r_R$, and $r_0$ (with $r_0= 10 \,\, \si{\micro\metre} $) in Table III are standard in MC literature. They model an MC channel where human insulin hormone is utilised as information molecules \cite{ISI_mitigating_methods_2015}. Please note that the RLL codes of order $i$ have the same order and normalized signal interval values with the corresponding RLIM codes of order $i$.

To determine the optimal static threshold  and scaling constant\footnote{ Dynamic detection algorithm assumes at least one 1-bit per code, which is the case for all RLIM$_i(n)$ but not necessarily for Uncoded, ISI-free, Hamming, and RLL schemes. To address this issue for these coding schemes, we used a modified dynamic detection algorithm from \cite{moac}, which introduces additional channel-specific parameters $min$ and $spacing$  alongside $a$. Through pilot signal transmissions for these schemes, we determined $min$, optimal $a$, and optimal $spacing$ values. For RLL codes, we adopted the same static spacing values (i.e. the $n$ values) as those used for the RLIM codes.} values for each coding scheme across different channel parameters, we transmitted a total of $53760$ bits by encoding $7$  different randomly chosen $7680$-bits sequences. Each of the $7$ runs was independent. For instance, RLIM$_1(24,16)$ encodes $7680$ bits using $11520$ bits, where all of the bits are transmitted in a contiguous manner. In MC simulations, imitating ISI, which stem from the accumulation of molecules over time, is important. Thus, uninterruptedly simulating long contiguous codewords with a high value of channel capacity (which we took as $200$) provides a realistic representation of the MC channel, as we do here.

\begin{figure}[t]
\centering
\captionsetup[subfigure]{labelformat=parens, labelsep=space}

\newlength{\colWone}   \setlength{\colWone}{0.495\columnwidth}
\newlength{\rowgapone} \setlength{\rowgapone}{0mm}
\setlength{\tabcolsep}{0pt}

\newlength{\LegendWone}    \setlength{\LegendWone}{35cm}      
\newlength{\LegendDropone} \setlength{\LegendDropone}{0mm}         
\newlength{\ImgHABone}     \setlength{\ImgHABone}{0.5\columnwidth}  
\newlength{\Hgapone}       \setlength{\Hgapone}{0.6mm}              

\captionsetup[subfigure]{skip=-3pt}
\captionsetup{skip=-2pt}

\begin{tabular}{@{}p{\colWone}@{\hspace{\Hgapone}}p{\colWone}@{}}

\multirow[t]{2}{\colWone}{%
  \parbox[t][\dimexpr 2\ImgHABone+\rowgapone\relax][t]{\colWone}{%
    \vspace*{\LegendDropone}%
    \centering
    \includegraphics[
      width=\LegendWone,
      height=\dimexpr 2\ImgHABone+\rowgapone-\LegendDropone\relax,
      keepaspectratio
    ]{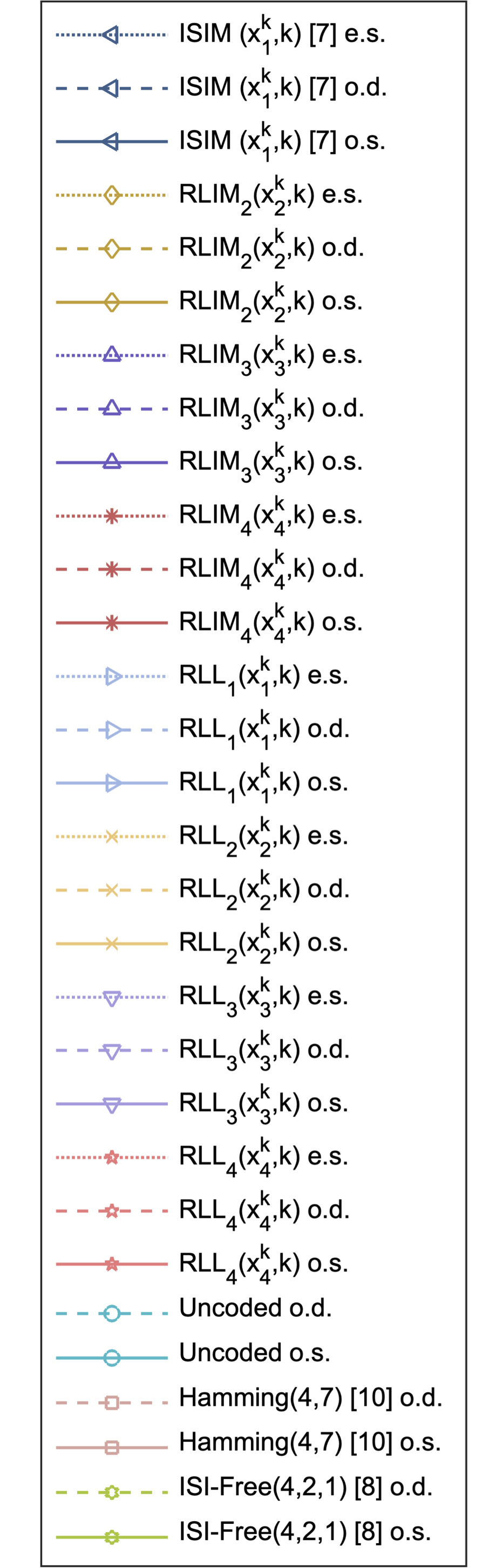}%
  }%
}
&

\hspace*{-1mm}%
\begin{subfigure}[t][\ImgHABone][c]{\colWone}\centering
  \includegraphics[width=\linewidth,height=\ImgHABone,keepaspectratio]{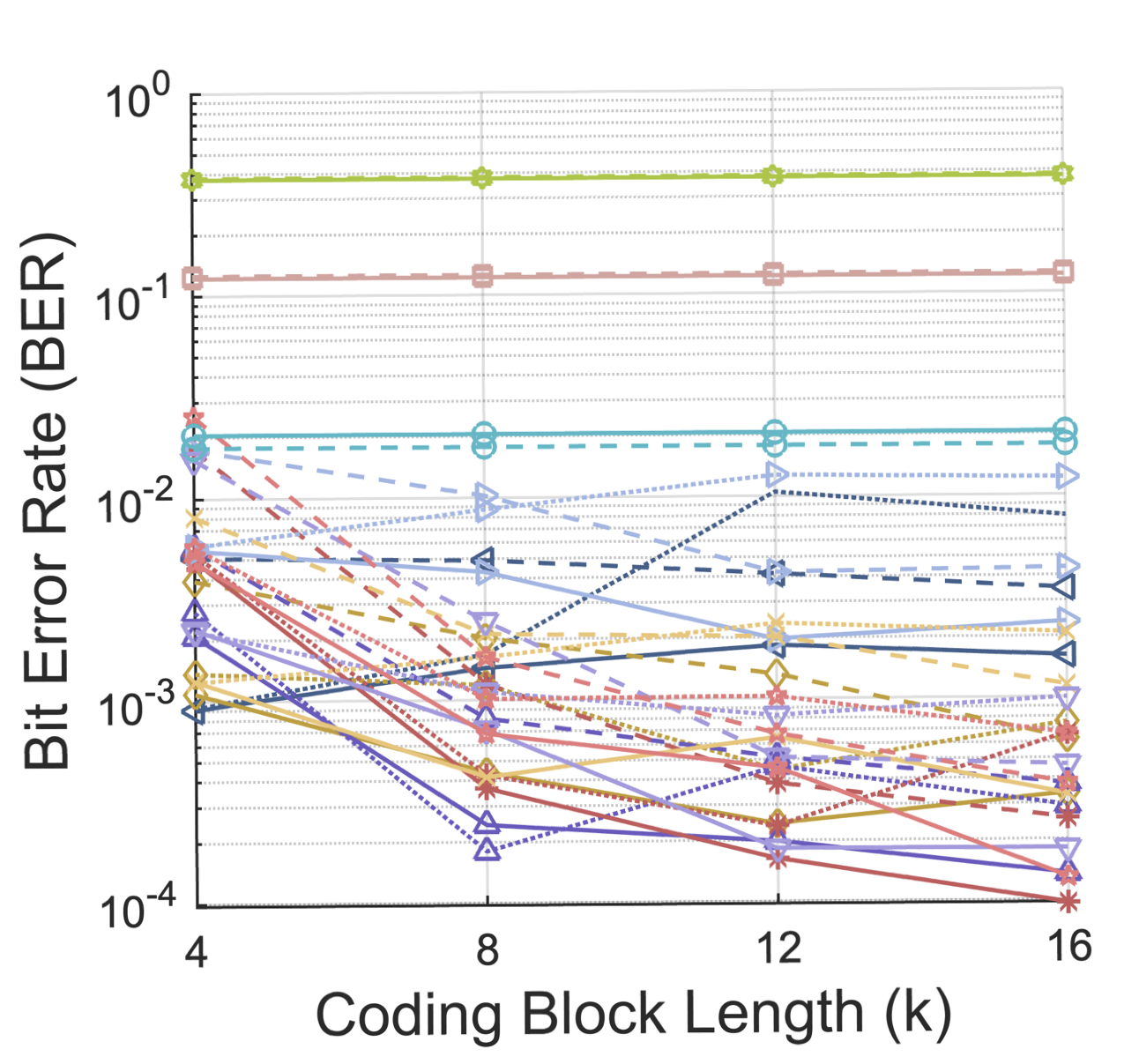}
  \vspace{-6pt}
  \caption{\scalebox{1}[1]{$t_s=200~\mathrm{ms},\ \hspace{-0.1cm} \sigma_n^2=0, \hspace{-0.1cm}\ r_0=10~\mu\mathrm{m}$}} \hspace{-1.7cm} \makebox[0.73\linewidth][r]{\scalebox{0.73}[0.73]{$M=800$}}
   \label{fig:one-a}
  \hspace{0.2cm}
\end{subfigure}
\\[1cm]

&

\hspace*{-1mm}%
\begin{subfigure}[t][\ImgHABone][c]{\colWone}\centering
  \includegraphics[width=\linewidth,height=\ImgHABone,keepaspectratio]{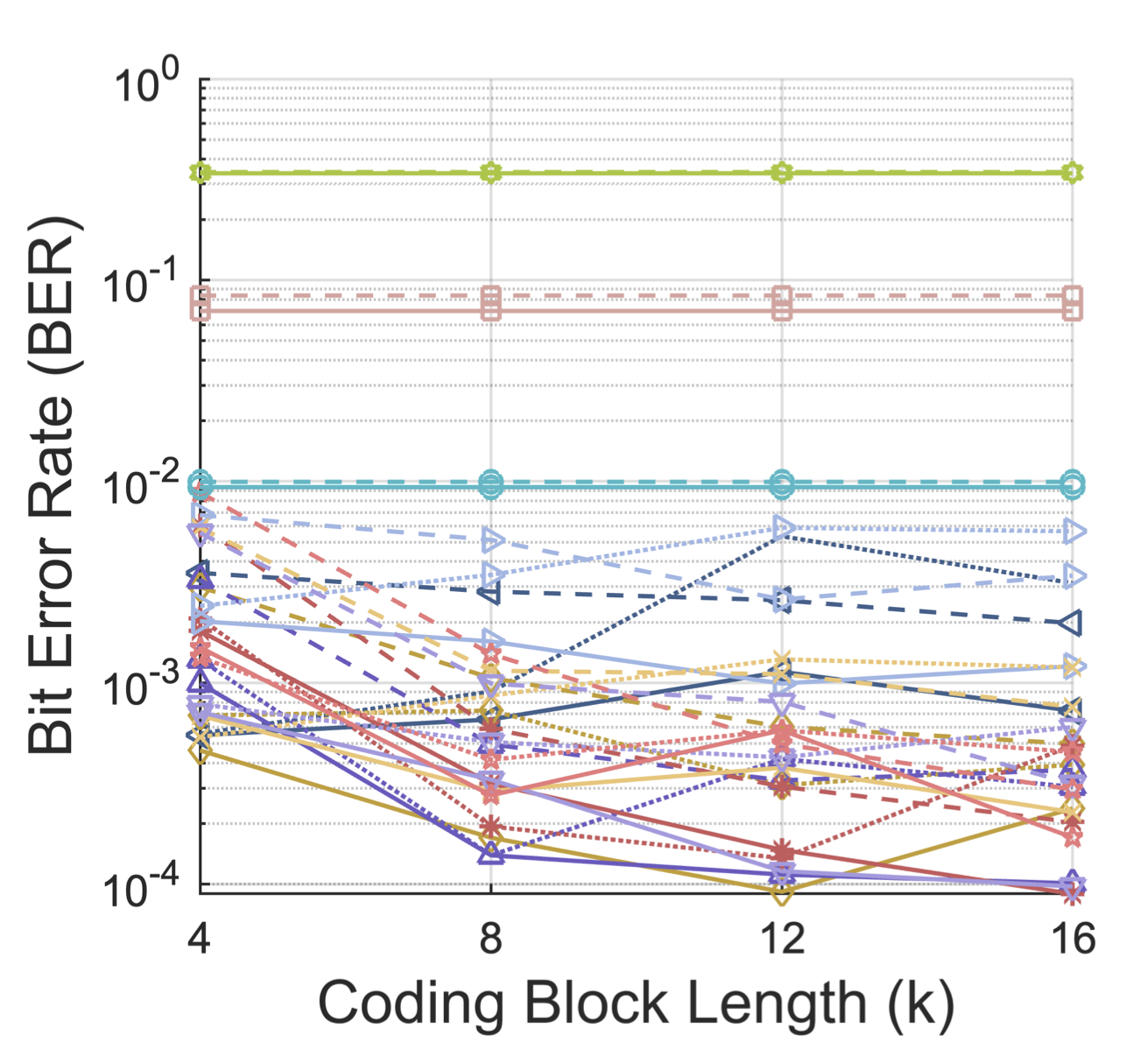}
  \vspace{-6pt}
  \caption{\scalebox{1}[1]{$t_s=250~\mathrm{ms},\ \hspace{-0.1cm} \sigma_n^2=0, \hspace{-0.1cm}\ r_0=10~\mu\mathrm{m}$}} \hspace{-1.7cm} \makebox[0.73\linewidth][r]{\scalebox{0.73}[0.73]{$M=400$}}
   \label{fig:one-a}
  \hspace{0.2cm}
\end{subfigure}
\\[1cm]

\end{tabular}

\vspace{16pt}
\caption{MC Simulation Results for Different Block Lengths. Abbreviations: “o.d.” = optimal dynamic, “o.s.” = optimal static, “e.s.” = estimated static.}
\vspace{-0pt}
\label{fig:onecol}
\end{figure}

  After having determined the optimal values of the static threshold and scaling constants (along with $min$ and optimal $spacing$ values for Uncoded, ISI-free, and Hamming schemes $^{5}$), we sent a total of $2257920$ bits ($7$ contiguous runs of the encodings of randomly chosen bit sequences of length $322560$) to calculate the respective BER values. When $t_s$=250 ms and $r_0$=9.5 \si{\micro\metre}, we multiplied the number of test bits by $2$ to obtain more accurate BER values. Please note that these specific number of bits have been chosen to ensure that all different block lengths of 4, 8, 12, and 16 perfectly divide these numbers, ensuring a fair comparison among coding methods in Fig. 4. Note that in Figs. 4-6, RLIM$_1(n,k)$ is referred to as ISI-mitigating codes \cite{best_channel_coding_2020} of length $n$ with block length $k$ (i.e., ISIM $(n,k)$), as they are equivalent \footnote{The ISIM code construction in \cite{best_channel_coding_2020} does not enforce the 1-bit minimization criterion or provide the full details of how subset selection is done.  Consequently, we assume ISIM \cite{best_channel_coding_2020} selects codewords by minimizing the number of 1-bits, which makes them equal to our RLIM codes of order 1.}.

\vspace{0.2cm}

\subsection{Evaluation of Simulation Results}

\subsubsection{Impact of Block Length on Performance}

\begin{table}[!t]
\caption{Values of $\{x^k_i\}$ in Figs. 4(a) and (b)}
\centering
\begin{tabular}{c|c c c c}
\hline
$\boldsymbol{k}$ & $\boldsymbol{x^k_1}$ & $\boldsymbol{x^k_2}$ & $\boldsymbol{x^k_3}$ & $\boldsymbol{x^k_4}$ \\
\hline
$\boldsymbol{4}$  & $7$  & $9$  & $11$ & $13$ \\
$\boldsymbol{8}$  & $13$ & $16$ & $20$ & $23$ \\
$\boldsymbol{12}$ & $18$ & $24$ & $28$ & $33$ \\
$\boldsymbol{16}$ & $24$ & $31$ & $37$ & $42$ \\
\hline
\end{tabular}
\end{table}

The simulation results for different block lengths $k$ (i.e., different information sets $\{0, 1\}^{k}$) are provided in Fig. 4(a) and (b). The corresponding values of $\{x^k_i\}$, whose derivations follow a similar process to those given in formulae 21-24, are shown in Table IV. These figures show that, as $k$ increases, the best attainable BER, defined as the minimum across all compared schemes at each $k$, decreases; for example, the best BER at $k=8$ is lower than at $k=4$. Notably, in both Figs. 4(a) and (b), the best overall performance among all coding schemes and block lengths is achieved when a block length of 16 is used. This is why we chose $k=16$ for the remainder of comparisons. Note that the normalizations of molecule count and signal interval values for block lengths 4, 8, and 12 (not shown for brevity) follow the same process as for block length 16, as detailed in Tables I and II.

\begin{figure}[t]
\centering
\captionsetup[subfigure]{labelformat=parens, labelsep=space}

\newlength{\colW}   \setlength{\colW}{0.495\columnwidth}
\newlength{\rowgap} \setlength{\rowgap}{0mm}
\setlength{\tabcolsep}{0pt}

\newlength{\LegendW}    \setlength{\LegendW}{20cm}          
\newlength{\LegendDrop} \setlength{\LegendDrop}{0mm}           
\newlength{\ImgHAB}     \setlength{\ImgHAB}{0.45\columnwidth}  
\newlength{\ImgHRest}   \setlength{\ImgHRest}{0.5\columnwidth}
\newlength{\Hgap}       \setlength{\Hgap}{0.6mm}              

\captionsetup[subfigure]{skip=-3pt}
\captionsetup{skip=-2pt}

\begin{tabular}{@{}p{\colW}@{\hspace{\Hgap}}p{\colW}@{}}

\multirow[t]{2}{\colW}{%
  \parbox[t][\dimexpr 2\ImgHAB+\rowgap\relax][t]{\colW}{%
    \vspace*{\LegendDrop}%
    \centering
    \includegraphics[
      width=\LegendW,
      height=\dimexpr 2\ImgHAB+\rowgap-\LegendDrop\relax,
      keepaspectratio
    ]{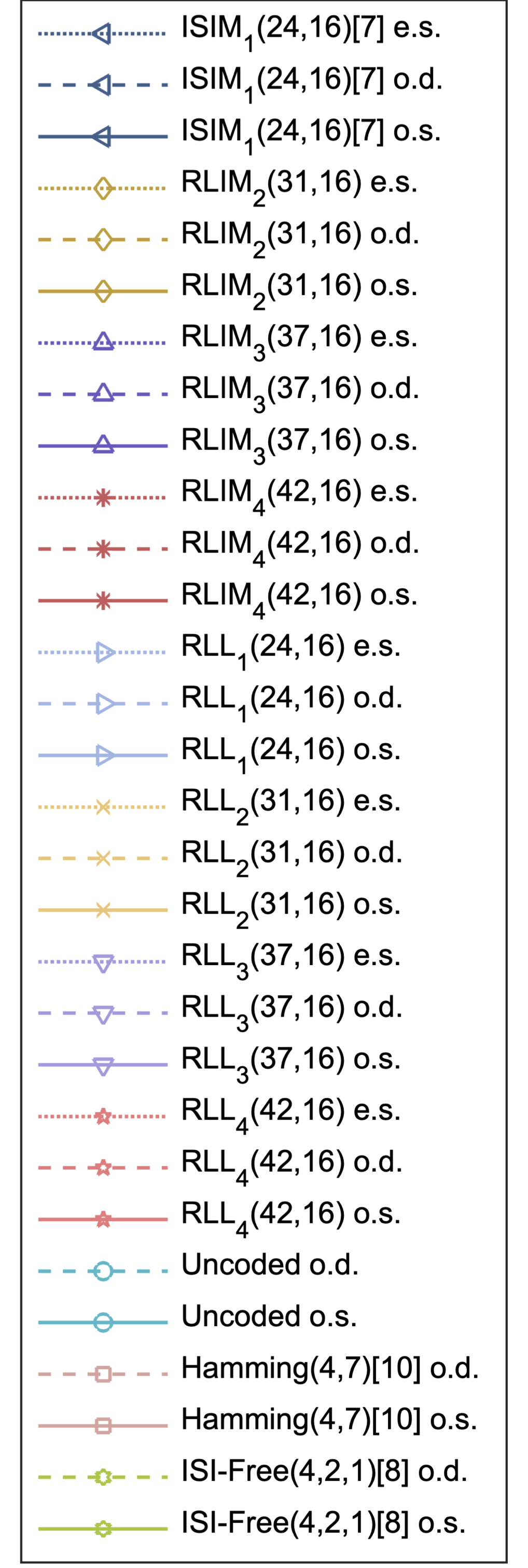}%
  }%
}
&

\hspace*{-1mm}%
\begin{subfigure}[t][\ImgHAB][c]{\colW}\centering
  \includegraphics[width=\linewidth,height=\ImgHAB,keepaspectratio]{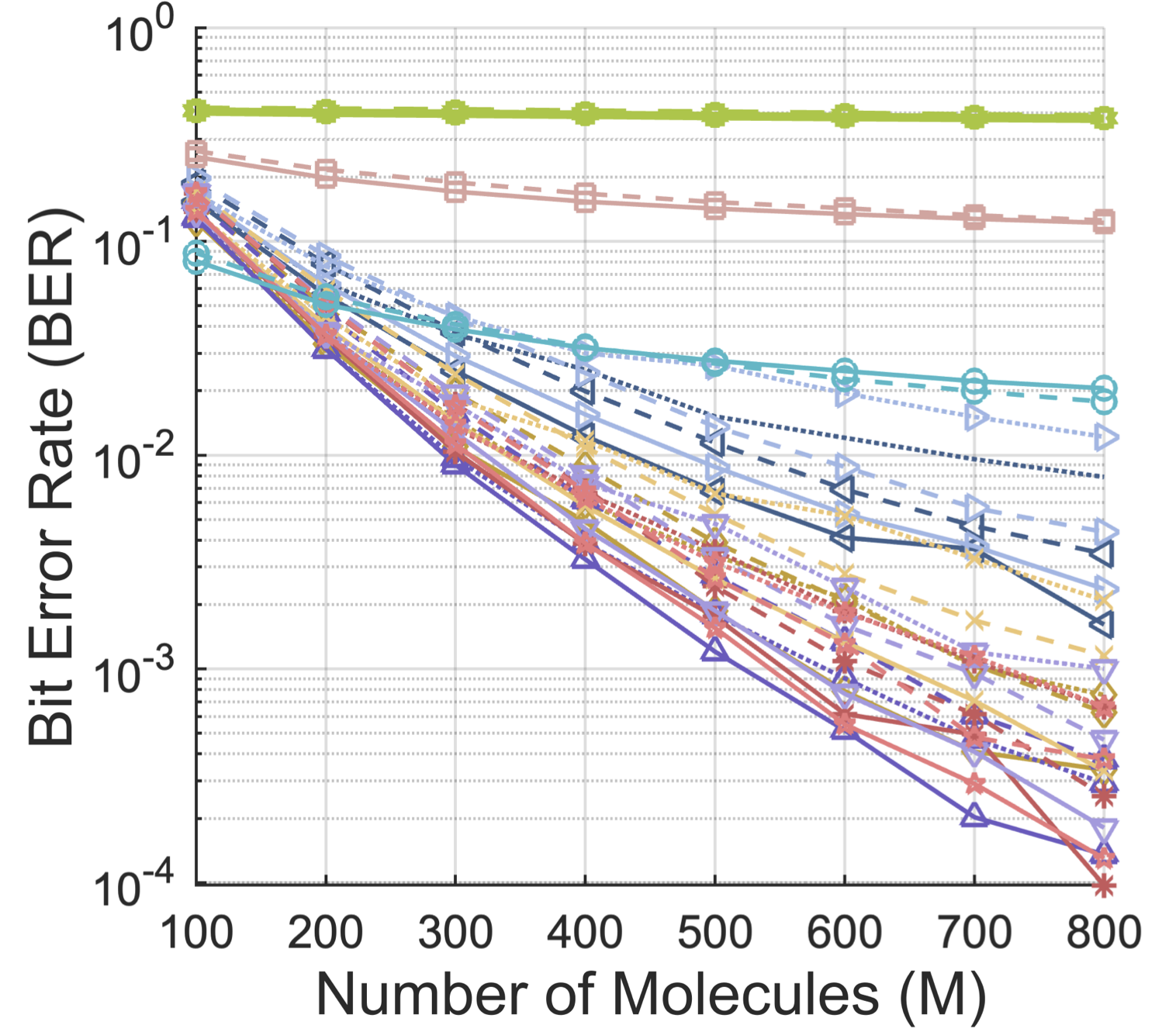}
  \vspace{-6pt}
  \caption{\scalebox{1}[1]{$t_s=200~\mathrm{ms},\ \hspace{-0.1cm} \sigma_n^2=0, \hspace{-0.1cm}\ r_0=10~\mu\mathrm{m}$}}\label{fig:a}
\end{subfigure}
\\[-5pt]

&

\hspace*{-1mm}%
\begin{subfigure}[t][\ImgHAB][c]{\colW}\centering
  \includegraphics[width=\linewidth,height=\ImgHAB,keepaspectratio]{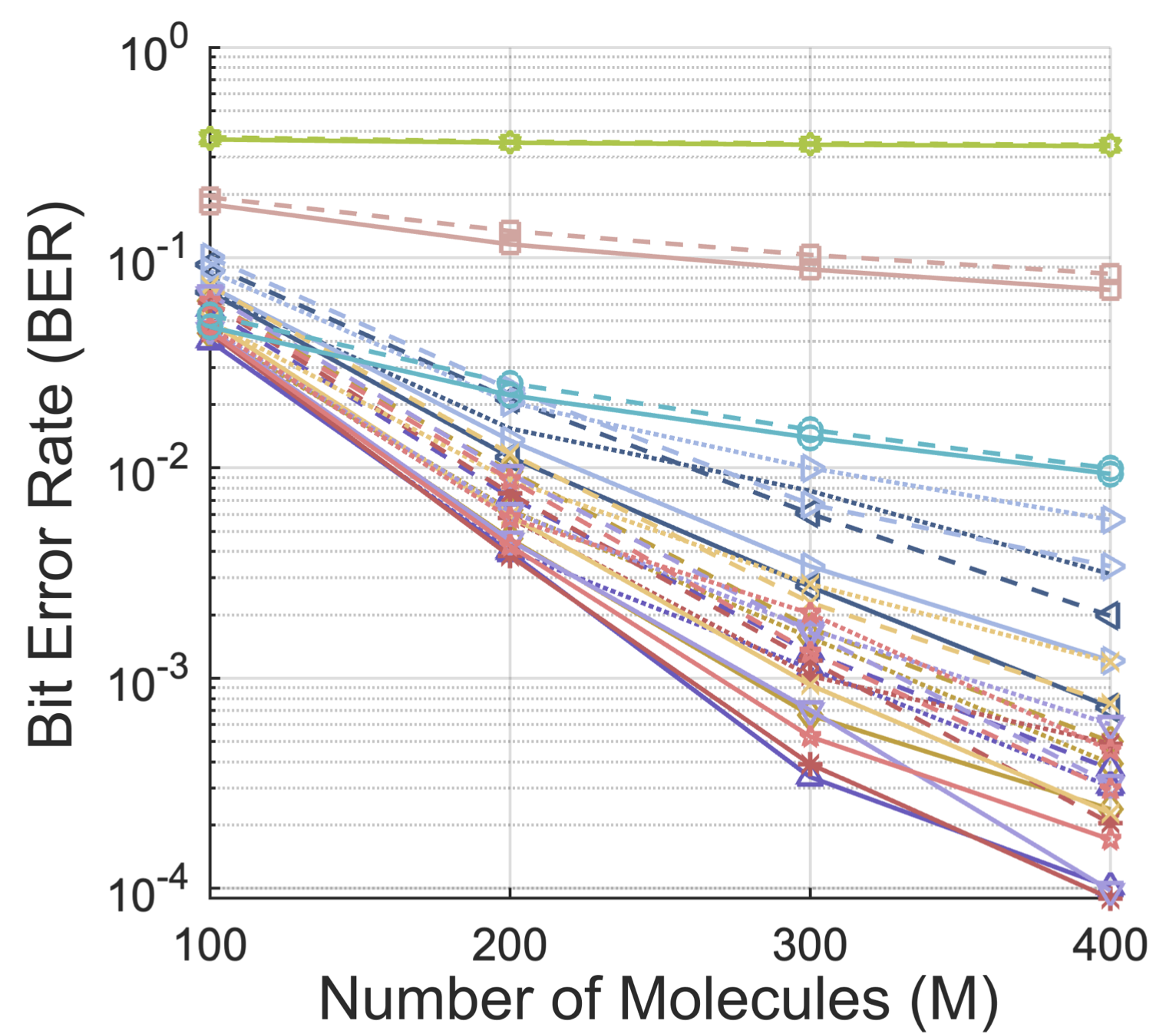}
  \vspace{-6pt}
  \caption{\scalebox{1}[1]{$t_s=250~\mathrm{ms},\ \hspace{-0.1cm} \sigma_n^2=0, \hspace{-0.1cm} \ r_0=10~\mu\mathrm{m}$}}\label{fig:b}
\end{subfigure}
\\[-6pt]

\begin{subfigure}[t][\ImgHRest][c]{\colW}\centering
  \includegraphics[width=\linewidth,height=\ImgHRest,keepaspectratio]{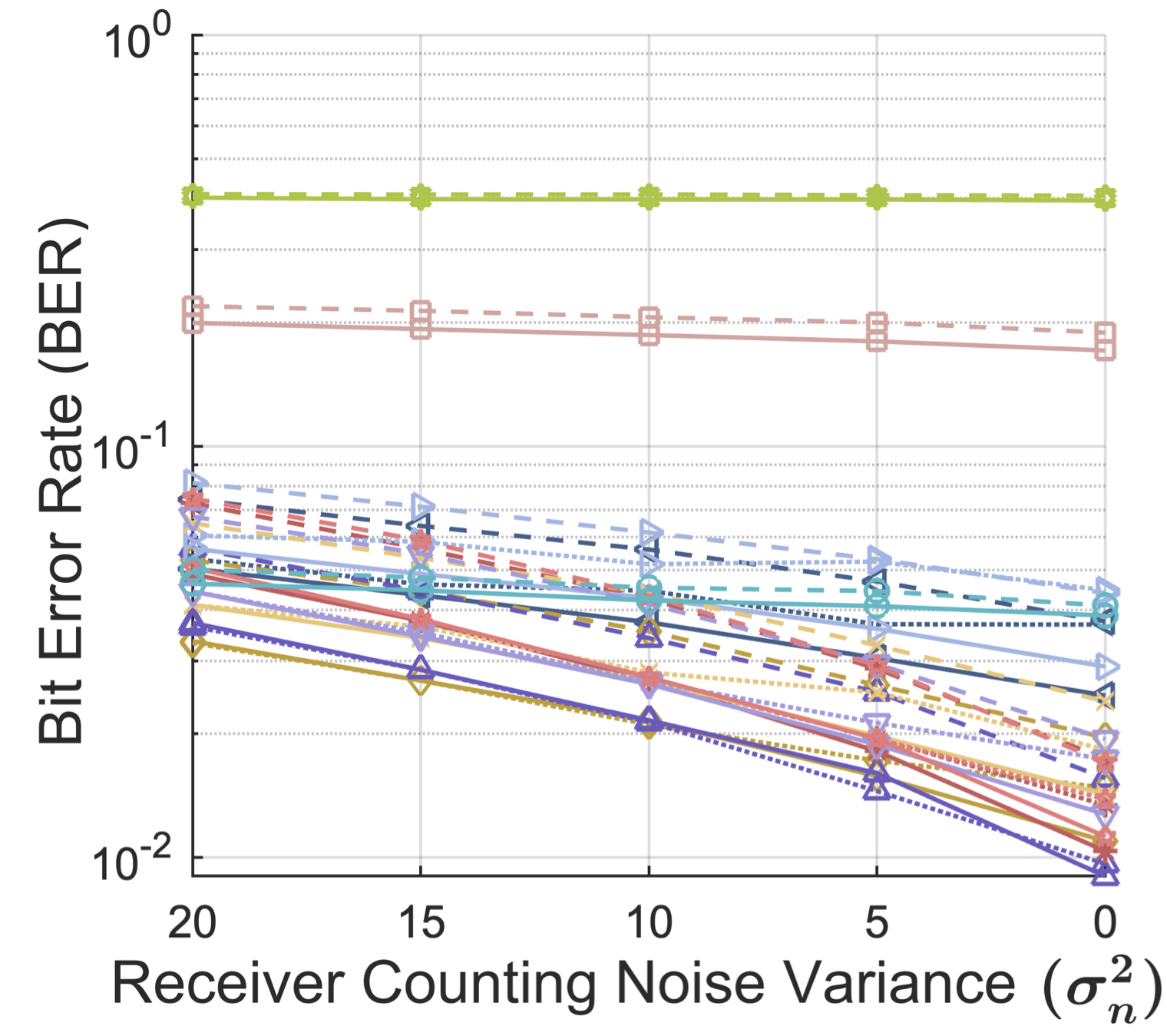}
  \vspace{-6pt}
  \caption{\scalebox{0.95}[1]{$t_s=200~\mathrm{ms}, \hspace{-0.1cm}\ M=300, \hspace{-0.1cm}\ r_0=10~\mu\mathrm{m}$}}\label{fig:c}
\end{subfigure}
&
\setcounter{subfigure}{4} 
\begin{subfigure}[t][\ImgHRest][c]{\colW}\centering
  \includegraphics[width=\linewidth,height=\ImgHRest,keepaspectratio]{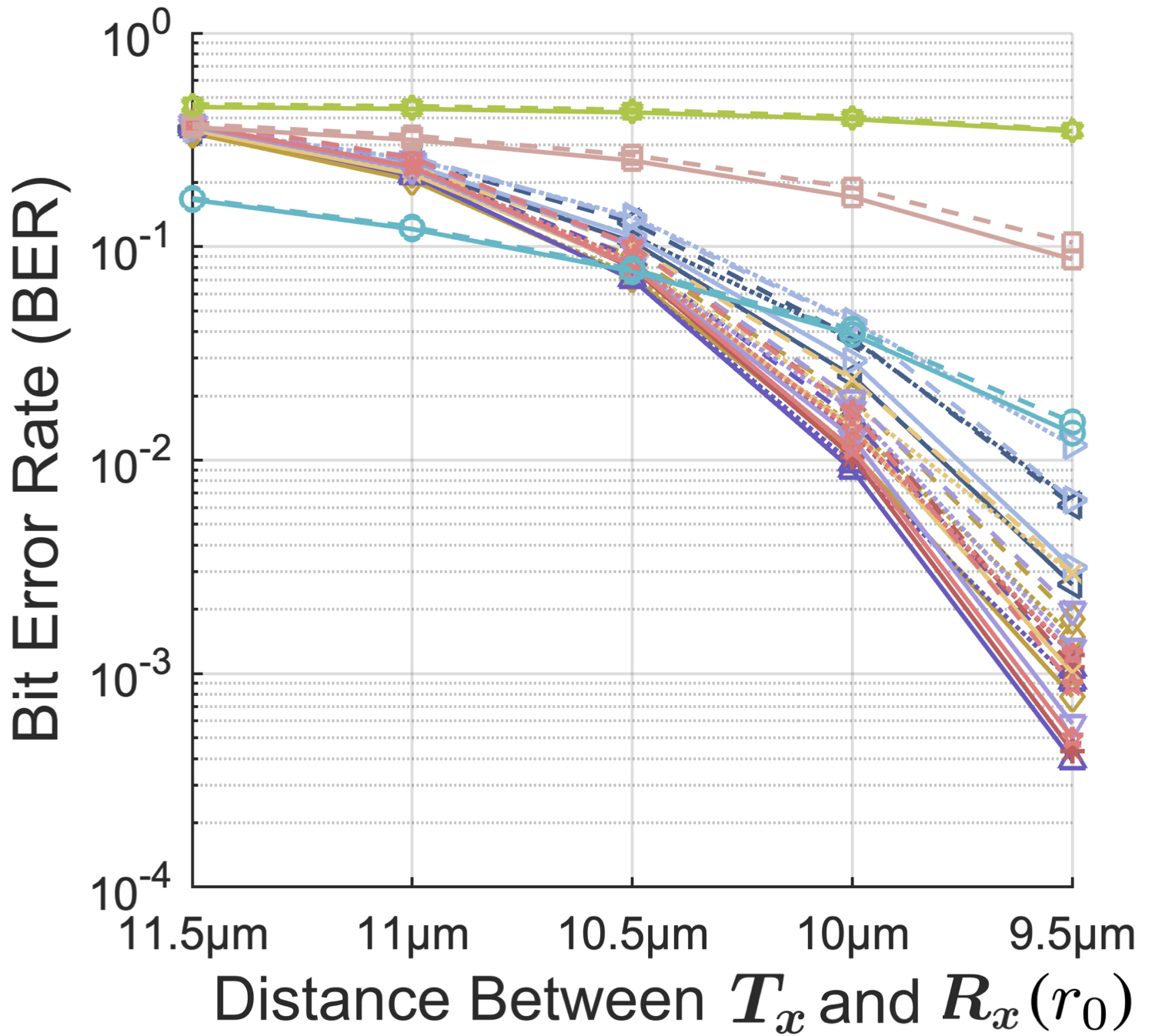}
  \vspace{-6pt}
  \caption{\scalebox{0.95}[1]{$t_s=200~\mathrm{ms}, \hspace{-0.1cm} \ \sigma_n^2=0, \hspace{-0.1cm}\ M=300$}}\label{fig:e}
\end{subfigure}
\\[-6pt]

\setcounter{subfigure}{3} 

\begin{subfigure}[t][\ImgHRest][c]{\colW}\centering
  \includegraphics[width=\linewidth,height=\ImgHRest,keepaspectratio]{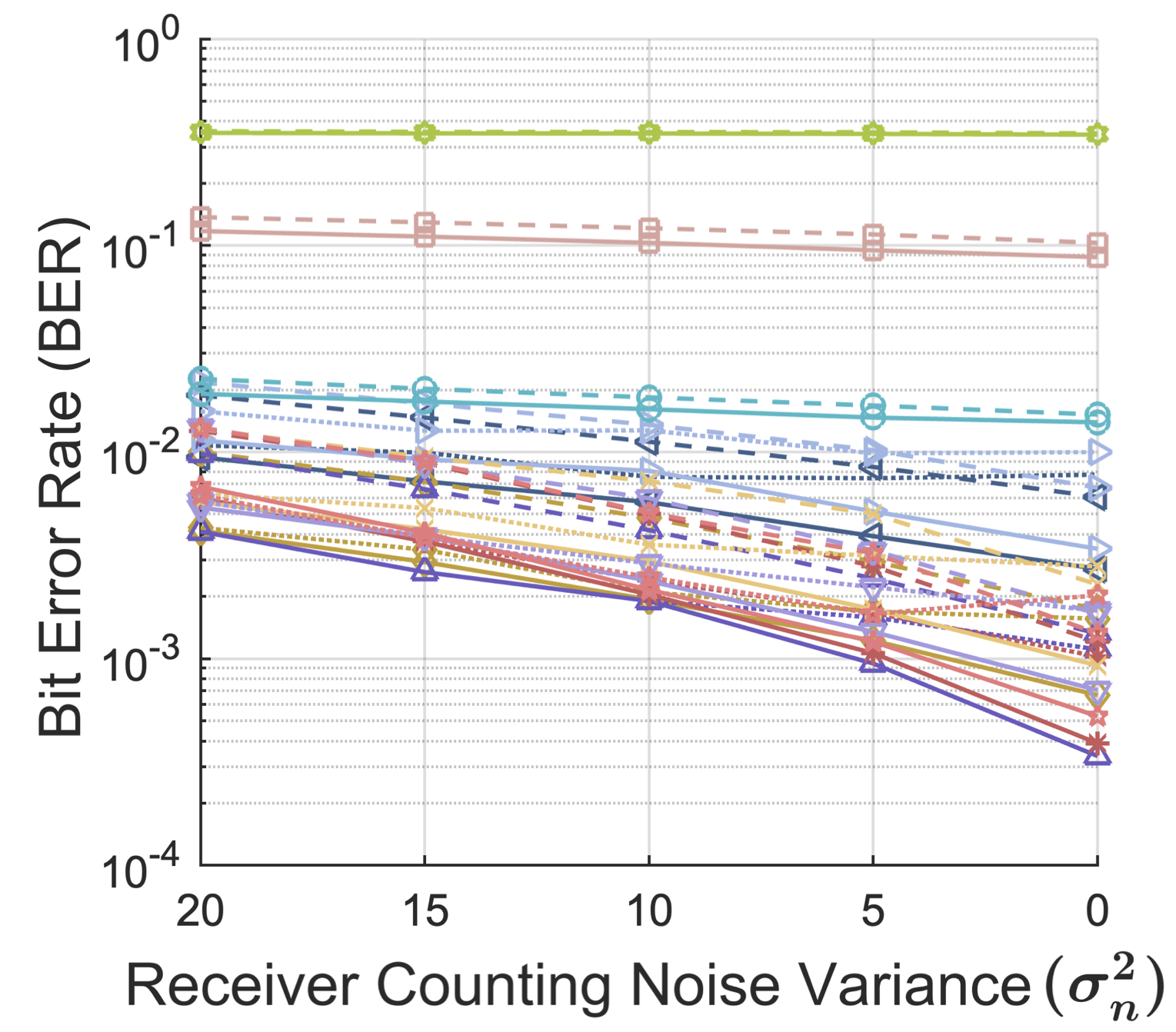}
  \vspace{-6pt}
  \caption{\scalebox{0.95}[1]{$t_s=250~\mathrm{ms}, \hspace{-0.1cm}\ M=300, \hspace{-0.1cm}\ r_0=10~\mu\mathrm{m}$}}\label{fig:d}
\end{subfigure}

\setcounter{subfigure}{5} 
&
\begin{subfigure}[t][\ImgHRest][c]{\colW}\centering
  \includegraphics[width=\linewidth,height=\ImgHRest,keepaspectratio]{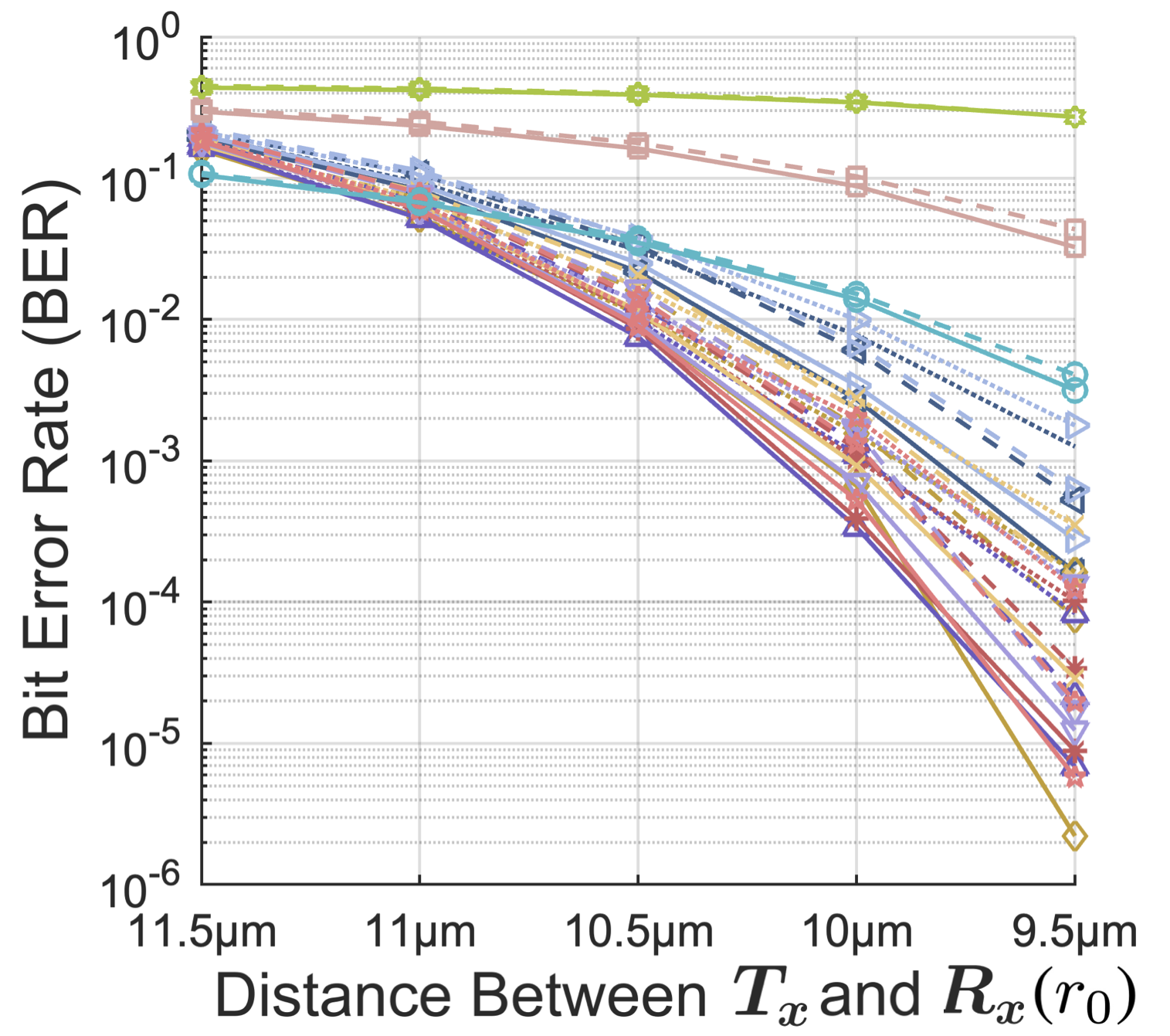}
  \vspace{-6pt}
  \caption{\scalebox{1}[1]{$t_s=250~\mathrm{ms},\ \sigma_n^2=0,\ M=300$}}\label{fig:f}
\end{subfigure}
\\[-2pt]

\end{tabular}

\vspace{8pt}
\caption{MC Simulation Results for Different Parameters}
\vspace{-10pt}
\label{fig:main}
\end{figure}

\begin{table}[!b]
\centering
\caption{Elementwise BER comparison of RLIM$_i$ vs.\ RLL$_i$.}
\label{tab:rlim-vs-rll-elementwise}
\renewcommand{\arraystretch}{1.05}
\setlength{\tabcolsep}{5pt}
\begin{tabular}{c|cc|cc|c}
\hline
$i$ & \multicolumn{2}{c|}{RLIM$_i$ wins} & \multicolumn{2}{c|}{RLL$_i$ wins} & ties \\
\cline{2-3}\cline{4-5}
 & count & mean $(\mathrm{RLL}/\mathrm{RLIM})$ & count & mean $(\mathrm{RLIM}/\mathrm{RLL})$ & \\
\hline
1 & 101 & $1.518\times$ & 1  & $1.143\times$ & 0 \\
2 & 97  & $1.669\times$ & 5  & $1.100\times$ & 0 \\
3 & 94  & $1.542\times$ & 7  & $1.260\times$ & 1 \\
4 & 78  & $1.321\times$ & 24 & $1.167\times$ & 0 \\
\hline
\end{tabular}
\end{table}

\subsubsection{Comparison Between RLL and RLIM}
As shown in Section III.B, Algorithm~2 (RLIM Error Correction) is equivalent to Viterbi with a last-wins tie-break and, on our MC channels, achieves substantially lower BER than hard-decision Viterbi with other common tie-breaks (e.g., first-wins and random). We now compare RLIM–corrected RLL directly against our RLIM codes, which additionally minimise the total number of 1–bits in the codebook (see Table~\ref{tbl_combined_parameters}); this weight minimisation increases the number of molecules available per transmitted \(1\)-bit under the normalisation in Section~\ref{sec:performance}, and therefore is expected to further reduce BER.

We perform an elementwise comparison across all distinct simulation parameters. For each order \(i\in\{1,2,3,4\}\) and each distinct data point, we mark a “win” for the method with the lower BER and record the BER fold ratio as the winner’s margin. The summary in Table V reports, for each \(i\), how many positions each method wins and the mean fold among those winning positions. Across all orders, RLIM wins \(370\) of \(408\) comparisons with a weighted mean improvement of \(\approx\!1.522\times\) at the winning points, whereas the rarer RLL wins average \(\approx\!1.175\times\). The only order with a noticeable number of RLL wins is \(i{=}4\) (24/102). This aligns with Table~\ref{tbl_combined_parameters}: at \(i{=}4\), RLIM and RLL have the closest 1-bit counts, yielding smaller normalisation gaps and, consequently, tighter BER margins. Overall, RLIM maintains a clear advantage over RLL.

\subsubsection{Effects of Different Parameters}

As can be inferred from Fig. 5, higher molecule count and signal interval values reduce BER, while higher $r_0$ and receiver noise variance $\sigma_n^2$ levels degrade it. As the noise variance increases, the advantage of RLL and RLIM codes over uncoded transmissions decreases. Increasing the coding order $i$ initially improves BER, but we conjecture that benefits diminish after a certain value of $i$, especially with shorter coding block lengths, as can be seen in Fig. 5(a) and (b). Additionally, at greater distances between T$_x$ and R$_x$ ($r_0$), the performance of all coding methods (including Uncoded) degrades significantly, becoming unreliably worse. In such situations, the Uncoded method, however, surpasses all others, as shown in Fig. 5(e) and (f).

\subsubsection{Storage Requirements and Trade-offs}

While increasing the block length $k$ can reduce BER, it exponentially raises memory requirements for RLIM codes, as each RLIM$_i(n,k)$ requires $n \cdot 2^k$ bits of storage space. These trade-offs  must  be  carefully considered  for practical  applications of  the proposed coding scheme. For $k=8, 12, 16$, the proposed RLIM codes of order $i \ge 2$ still outperform others (in their respective detection technique categories), as demonstrated in Fig. 4(a) and (b). Accordingly, if data storage capacity is limited, opting for $k=8$ or $k=12$ instead of $k=16$ is advisable. For future comparative simulations of our proposed family of codes, we recommend using a coding block length of at least $12$ for a fair comparison, with $k=16$ being a more preferable option.

\subsubsection{Constraint-Level Coincidence with $(i,\infty)$–RLL in Drift-Free Channels}
For a non-drift channel, a key observation is that for all RLIM, the optimal static threshold detection method consistently outperforms the optimal dynamic threshold method. Consequently, dynamic threshold detection would only be necessary when the parameters of MC channel fluctuates over time. Henceforth, it is no longer necessary to assume the presence of a single 1-bit in a code. Thus, for future uses with static threshold through a non-drift static channel, we make the enhancement in (25) for RLIM codes, so the constraint set coincides with the classical run-length-limited (RLL) codes of order ($i,\infty$) of length $n$ \cite{RLL} with leading merging 0-bits. Please note that implementing this enhancement would eliminate the need for Lines 7–10 in Algorithm 1. The critical minimization of 1-bits property of practical RLIM$_i(n,k)$ codes still preserves their distinct character and makes them advantageous over RLIM-corrected classical RLL codes as shown in the simulation results.

\begin{equation}
\widehat{\mathrm{RLIM}}_{i}(n)
   \;=\;
   \begin{bmatrix}
      \mathbf 0^{\,i}_{|C_i(n-i)|} & C_i(n-i)
   \end{bmatrix}
   \;=\;
   (i,\infty)-\mathrm{RLL} (n)
\tag{25}
\end{equation}

\begin{tikzpicture}[overlay, remember picture]

  \draw[dashed, line width=0.5pt] (3.6, 0.6) -- (3.6, 1.4);
\end{tikzpicture}

\subsection{Time-Varying Drift Channel Model and Gains from Dynamic Threshold}
\label{subsec:drift_dynamic_threshold}

\begin{table}[!t]
\begin{center}
\caption{Drift and Particle-Tracking Parameters for Time-Varying MC Simulations}
\renewcommand{\arraystretch}{1.0}

\label{tbl_drift_params}
\begin{tabular}{p{6.5cm} l}
\hline
\bfseries{Parameter} & \bfseries{Value} \vspace{0.05cm}\\
\hline
Time step (\(\Delta t\))             & 1\,ms \\
Drift correlation timescale (\(\tau_{\text{drift}}\))    & 10\,s \\
Drift standard deviation (\(\sigma_v\))  & 10\,\si{\micro\metre/\second}\\
Mean drift velocity (\(\mathbf{v}_{\text{mean}}\))       & [1, 0, 0]\,\si{\micro\metre/\second}\\
Max molecule age  & 5\,s \\
Emitter (T$_x$) location                & [0, 0, 0]\,\si{\micro\metre} \\
Receiver (R$_x$) center                 & [\(r_0\), 0, 0]\,\si{\micro\metre} \\
\hline
\end{tabular}
\end{center}

\end{table}

In many real-world MC scenarios, the channel may exhibit random or time-varying drift, causing the arrival profiles of information molecules to fluctuate from one symbol interval to the next. Under such non-stationary conditions, a well-tuned dynamic threshold scheme can significantly outperform an optimal static threshold, since the latter fails to adapt to shifting channel conditions. To illustrate this advantage, we adopt a particle-tracking model, based on the simulator \cite{simulation}, that extends the discrete-time update formula given in (20).

At each simulation time step of duration \(\Delta t\), the drift velocity \(\mathbf{v}(t)\) is updated according to the Euler–Maruyama discrete-time approximation of an Ornstein--Uhlenbeck (OU) process with a nonzero mean:

\begin{equation}
\label{eq:OU_discretized}
\mathbf{v}(t+\Delta t) 
\hspace{-0.1cm}\;=\; \hspace{-0.1cm}
\mathbf{v}(t) 
\hspace{-0.1cm}\;+\; \hspace{-0.1cm}
\frac{1}{\tau_{\text{drift}}}\Bigl(\mathbf{v}_{\text{mean}}\hspace{-0.1cm}-\hspace{-0.1cm}\mathbf{v}(t)\Bigr)\Delta t 
\hspace{-0.1cm}\;+\; \hspace{-0.1cm}
\sqrt{\frac{2\,\sigma_v^2\,\Delta t}{\tau_{\text{drift}}}}\;\boldsymbol{\xi}(t),
\tag{26}
\end{equation}where \(\tau_{\text{drift}}\) is the drift correlation timescale, \(\sigma_v^2\) governs the variability of the drift, \(\mathbf{v}_{\text{mean}}\) is the mean drift velocity, and \(\boldsymbol{\xi}(t)\) is a three-dimensional vector whose components are independently drawn from \(\mathcal{N}(0,1)\). The Ornstein–Uhlenbeck process is a classical stochastic model widely employed to capture fluctuating physical quantities in many natural and engineered systems (see, e.g., \cite{gardiner2009stochastic} and \cite{vankampen2007stochastic}).
 Once the drift velocity is updated via \eqref{eq:OU_discretized}, each molecule’s position is updated by

\begin{equation}
\label{eq:position_update}
\mathbf{r}(t+\Delta t) 
\;=\; 
\mathbf{r}(t) 
\;+\; 
\mathbf{v}(t+\Delta t)\,\Delta t 
\;+\; 
\Delta \mathbf{w}(t),
\tag{27}
\end{equation}where \(\Delta \mathbf{w}(t)\) is a three-dimensional Gaussian random vector. Each component of \(\Delta \mathbf{w}(t)\) is drawn independently from \(\mathcal{N}\bigl(0,\,2D\,\Delta t\bigr)\), thereby modeling the diffusive displacement in accordance with (20) \footnote{An Ornstein-Uhlenbeck (OU) drift is colored, stationary, Gaussian, and Markovian, capturing finite-memory correlations. RLIM's dynamic-threshold gains, driven by adaptation to time-varying means, are expected to persist under finite-memory colored drifts; under long-memory or heavy-tailed drifts the benefit may diminish, and a comprehensive study of these regimes is an important direction for future work.}. See  \cite{Sun2015CapillaryCapacity} for capacity analysis for MC with blood-flow drift.

\begin{figure}[t]
\centering
\captionsetup[subfigure]{labelformat=parens, labelsep=space}

\newlength{\colWtwo}   \setlength{\colWtwo}{0.495\columnwidth} 
\newlength{\rowgaptwo} \setlength{\rowgaptwo}{0mm}
\setlength{\tabcolsep}{0pt}

\newlength{\LegendWtwo}    \setlength{\LegendWtwo}{\colWtwo}     
\newlength{\LegendDroptwo} \setlength{\LegendDroptwo}{15mm}             
\newlength{\ImgHtwo}       \setlength{\ImgHtwo}{0.42\columnwidth}    \newlength{\Hgaptwo}       \setlength{\Hgaptwo}{0.6mm}               

\captionsetup[subfigure]{skip=-3pt}
\captionsetup{skip=-2pt}

\begin{tabular}{@{}p{\colWtwo}@{\hspace{\Hgaptwo}}p{\colWtwo}@{}}

\multirow[t]{3}{\colWtwo}{%
  \parbox[t][\dimexpr 3\ImgHtwo+2\rowgaptwo\relax][t]{\colWtwo}{%
    \vspace*{\LegendDroptwo}%
    \centering
    \includegraphics[
      width=2.68cm,
      height=\dimexpr 3\ImgHtwo+2\rowgaptwo-\LegendDroptwo\relax,
      keepaspectratio
    ]{1.png}
  }%
}
&
\begin{subfigure}[t][\ImgHtwo][c]{\colWtwo}\centering
  \includegraphics[width=\linewidth,height=\ImgHtwo,keepaspectratio]{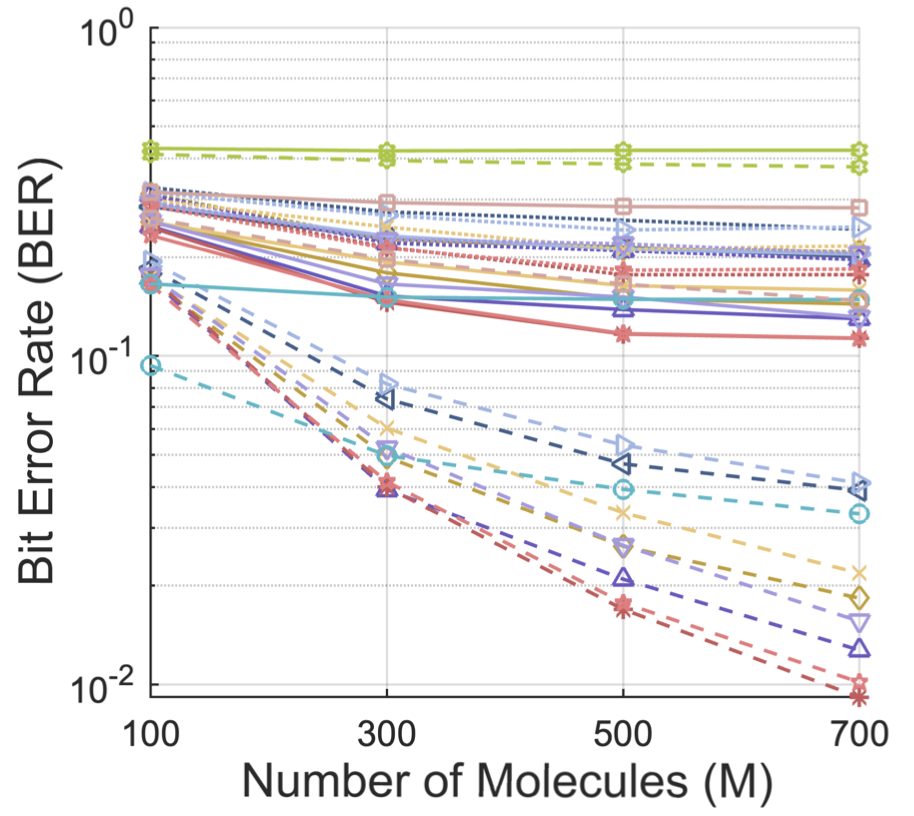}
  \vspace{6pt}
  \caption{$t_s=200~\mathrm{ms},\ \hspace{-0.1cm} \sigma_n^2=0, \hspace{-0.1cm}\ r_0=10~\mu\mathrm{m}$}\label{fig:two-a}
\end{subfigure}
\\[0pt]

&
\begin{subfigure}[t][\ImgHtwo][c]{\colWtwo}\centering
  \includegraphics[width=\linewidth,height=\ImgHtwo,keepaspectratio]{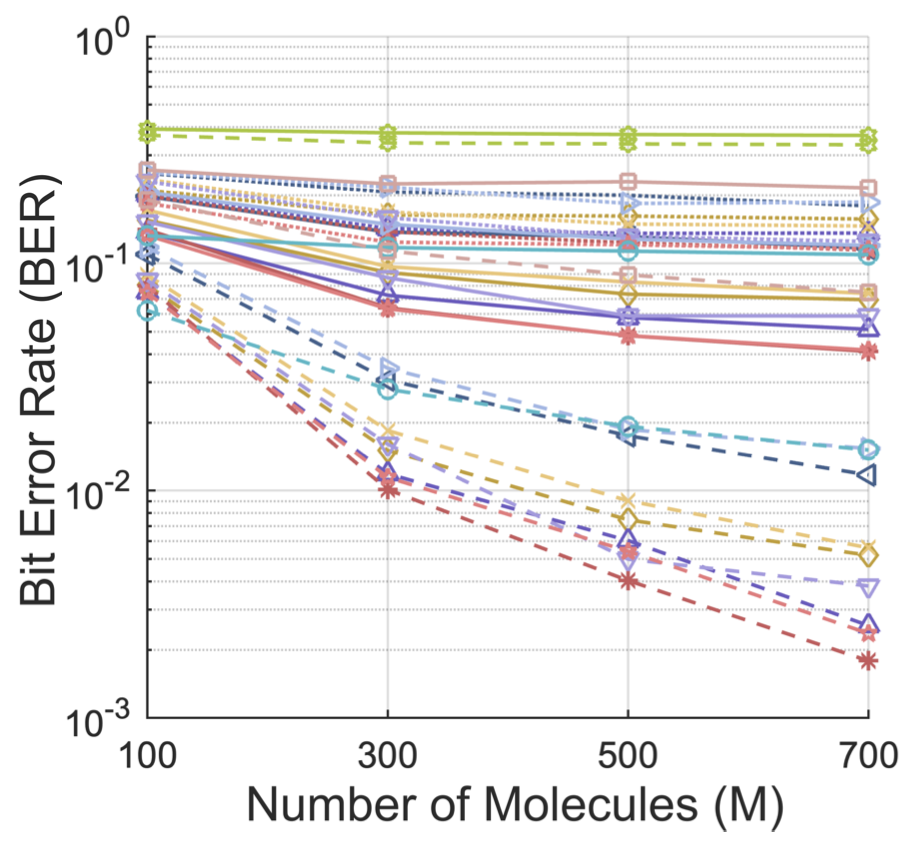}
  \vspace{6pt}
  \caption{$t_s=250~\mathrm{ms},\ \hspace{-0.1cm} \sigma_n^2=0, \hspace{-0.1cm}\ r_0=10~\mu\mathrm{m}$}\label{fig:two-b}
\end{subfigure}
\\[0pt]

&
\begin{subfigure}[t][\ImgHtwo][c]{\colWtwo}\centering
  \includegraphics[width=\linewidth,height=\ImgHtwo,keepaspectratio]{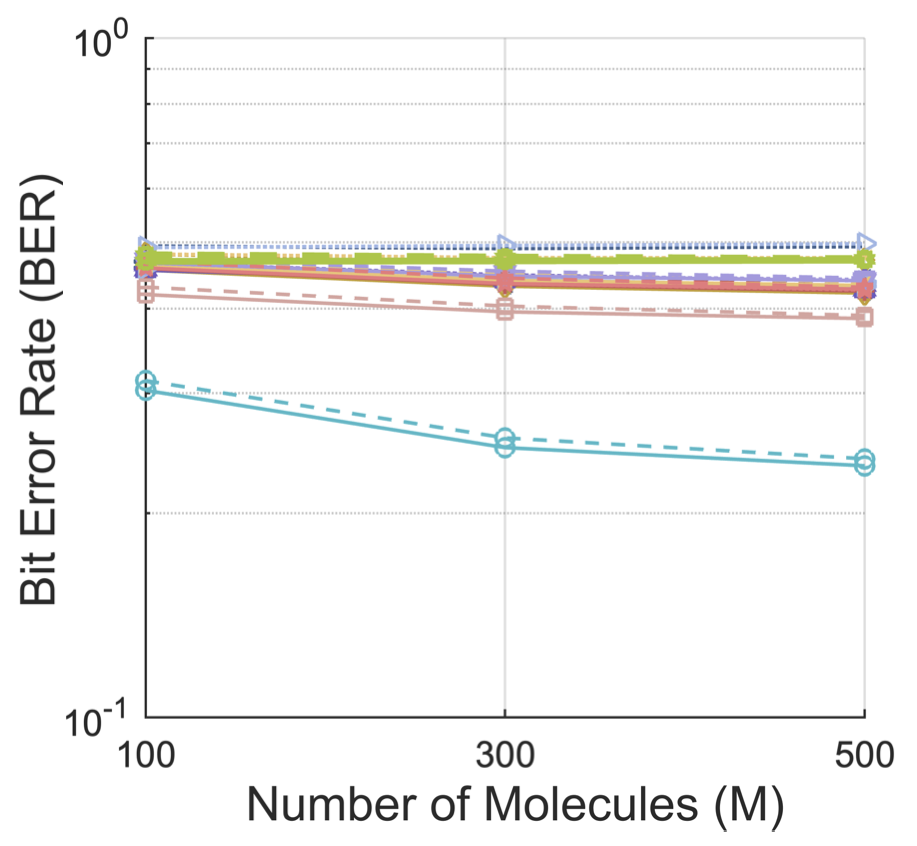}
  \vspace{6pt}
  \caption{$t_s=200~\mathrm{ms},\ \hspace{-0.1cm} \sigma_n^2=0, \hspace{-0.1cm}\ r_0=10~\mu\mathrm{m}$}\label{fig:two-c}
\end{subfigure}
\\[2pt]

\end{tabular}

\vspace{6pt}
\caption{Particle-Tracking MC Simulation Results with (a) and (b) denoting a time-varying OU drift channel and (c) denoting a channel with a transparent receiver}
\vspace{-2pt}
\label{fig:two}
\end{figure}

We use the same simulation parameters from Table III. Table VI further specifies the drift and particle-tracking parameters used in our drift-included MC simulations. The settings in Table VI produce a moderate but slowly varying drift that causes noticeable changes in the mean molecule arrival count over successive intervals. To prevent computational overload, the channel memory is set to 5~seconds. For a signal interval of 0.2~seconds, this corresponds to a channel memory of 25 in the Uncoded case. Since we have a time step of 1 ms, interval values given in Table I are rounded to the nearest integer.

The corresponding BER values of compared coding schemes for a generic MC channel with a time-varying drift are provided in Fig. 6 (a) and (b). Under these non-stationary conditions, a dynamic threshold consistently achieves lower bit error rates\footnote{For particle-tracking simulations, to determine the optimal detection parameters, we initially sent a total of $53760$ bits ($7$ contiguous runs of the encodings of randomly chosen bit sequences of length $7680$). After having determined the optimal detection parameters, to calculate each respective BER value, we then sent a total of $112000$ bits ($7$ contiguous runs of the encodings of randomly chosen bit sequences of length $16000$).} than the corresponding fixed threshold-based method, demonstrating the need for adaptive detection in such MC channels. As can be seen in Fig. 6 (a) and (b), RLIM and RLL codes with dynamic detection surpass the compared methods, except at $M=100$, where uncoded transmission obtain the lowest BER. In an elementwise comparison between RLIM and the RLIM-corrected RLL baseline across 96 distinct drifted settings, RLIM wins $72/96$ cases ($75\%$) with a mean fold improvement of $\approx 1.1\times$ at its winning points, whereas RLL wins $24/96$ with a mean fold of $\approx 1.056\times$. By order: $i{=}1$: $18$ vs.\ $6$ ($1.069\times$ vs.\ $1.042\times$); $i{=}2$: $20$ vs.\ $4$ ($1.113\times$ vs.\ $1.051\times$); $i{=}3$: $20$ vs.\ $4$ ($1.146\times$ vs.\ $1.095\times$); $i{=}4$: $14$ vs.\ $10$ ($1.080\times$ vs.\ $1.051\times$), with the margin narrowing at $i{=}4$ because the two codebooks have very similar 1-bit counts (hence smaller normalization gaps in Table II). These results, together with the fact that the dynamic detector of RLIM avoids the extra $min$ constant required by the dynamic detector for our RLL baseline, justify including at least one $1$-bit in each RLIM code under time-varying channels.

\vspace{0.1cm}

\subsection{Transparent Receiver Model}
\label{subsec:passive}

A transparent (passive) receiver is non-absorbing: molecules are not removed upon contact, and we model the reception by counting the molecules present inside the spherical receiver volume at the end of each signal interval \cite{passive}. The channel is drift-free; molecule positions are updated only by Brownian motion with $\Delta t=1$\,ms, and we use the same finite channel-memory cutoff, normalization, and number of training/test bits as in the OU particle-tracking experiments. Because molecules persist, ISI is stronger. As can be seen in Fig. 6 (c) Under this passive model at $t_s=0.2$\,s and $M\in\{100,300,500\}$, the increased ISI together with the shorter $t_s$ required for coded schemes outweighs their ISI-mitigation benefits, so the Uncoded baseline attains the lowest BER overall, although reliability remains poor (BER $>10^{-1}$).

\vspace{0.2cm}

\section{Conclusion}
\label{sec:conclusion}

We introduced RLIM, an $(i,\infty)$ run-length family that forms fixed-size codebooks by minimizing 1-bits, thereby increasing the per-symbol molecule budget under standard normalizations. We developed an optimal linear-time greedy decoder that targets ISI and is Viterbi-equivalent under a last-wins tie-break, while running at lower complexity. Extensive binomial and particle-tracking simulations show that RLIM attains lower BER than classical RLL and other prominent schemes across broad parameter ranges. For passive (non-absorbing) receivers, stronger ISI can narrow gains and favor uncoded at short $t_s$, with overall reliability limited. Future work should focus on developing more precise theoretical distributions for \({}^iN_{\hat{0}}^{\text{average}}\) and \({}^iN_{1}^{\text{average}}\), which could improve the BER of static threshold estimation and potentially eliminate the necessity of sending pilot signals before communication. Overall, the success of the proposed RLIM\(_i(n,k)\) codes represents a significant advancement in MC channel coding.

\appendices

\section{Derivation of (19)}
\label{app:derivation19}

Let $Q(x)=\tfrac{1}{\sqrt{2\pi}}\!\int_x^\infty e^{-t^2/2}\,dt$,
$\phi(x)=\tfrac{1}{\sqrt{2\pi}}e^{-x^2/2}$. Starting from (18),

\begin{equation}
\scalebox{1.0}{$
P(\tau_{static})
= P_{\hat{0}}\!\left[1 - Q\!\left(\frac{\tau_{static}-A}{\sqrt{B}}\right)\right]
+ P_1\, Q\!\left(\frac{\tau_{static}-C}{\sqrt{D}}\right).
$}
\tag{A.1}
\end{equation}

Taking the derivative gives
\begin{equation}
\scalebox{1.0}{$
\frac{dP}{d\tau_{static}}
= \frac{P_{\hat{0}}}{\sqrt{B}}\,\phi\!\left(\frac{\tau_{static}-A}{\sqrt{B}}\right)
- \frac{P_1}{\sqrt{D}}\,\phi\!\left(\frac{\tau_{static}-C}{\sqrt{D}}\right)=0
$}
\tag{A.2}
\end{equation}

Define
\[
\scalebox{1.0}{$
L := \log_e\!\left(\frac{\sqrt{D}\,P_{\hat{0}}}{\sqrt{B}\,P_1}\right),
$}\tag{A.3}\]

Using (A.2), substitute $\phi(x)=\tfrac{1}{\sqrt{2\pi}}e^{-x^2/2}$, cancel the common constants, rearrange, and apply $\log_e$ to both sides to obtain
\begin{equation}
\scalebox{1.0}{$
\frac{(\tau_{static}-A)^2}{2B} - \frac{(\tau_{static}-C)^2}{2D} = L
$}
\tag{A.4}
\end{equation}

Multiply by $2BD$ and expand:
\begin{equation}
\scalebox{0.85}{$
\begin{aligned}
(D-B)\,\tau_{static}^2
&\;+\; 2(BC-AD)\,\tau_{static} \\
&\;+\; \bigl(DA^2 - BC^2 - 2LBD\bigr) \;=\; 0
\end{aligned}
$}
\tag{A.5}
\end{equation}

Solving (A.5) gives
\begin{equation}
\scalebox{1.0}{$
\tau_{static}
= \frac{DA-BC}{D-B}
 \;\pm\; \frac{\sqrt{\,BD\bigl((C-A)^2 - 2(B-D)L\bigr)}}{D-B}.
$}
\tag{A.6}
\end{equation}

Choosing the physically relevant root (between the means, typically $A<C$) yields the (19).

\vspace{0.25cm}

\section*{Acknowledgment}

This work is dedicated to the memory of Muzaffer Şahin, the first author’s late grandfather, whose struggle with cancer motivated this research in Molecular Communication, with the recognition that fundamental progress in this field can support the development of more effective cancer therapies.

\vspace{0.25cm}

\section*{Code Availability}
 Python implementations of binomial simulator, code space generation, encoding, decoding, analytical threshold estimation, detection, optimal static threshold finder, and error correction algorithms for any chosen RLIM$_i(n,k)$ are provided in the following link:
 https://github.com/MelihSahinEdu/McChannelCoding.git

\bibliographystyle{IEEEtran}

\bibliography{References}

\begin{IEEEbiography}
[{\includegraphics[width=1in,height=1.25in,clip,keepaspectratio]{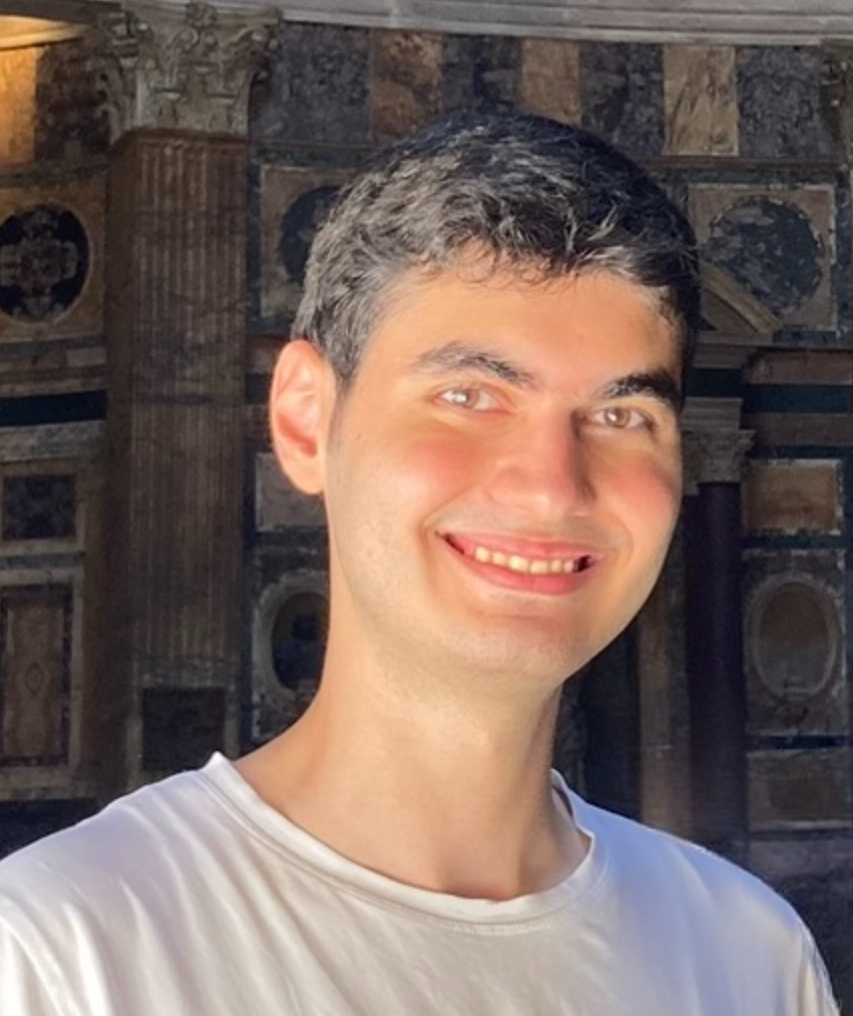}}]{Melih Şahin (Student Member, IEEE)}
is a PhD student in the Department of Engineering at the University of Cambridge. He received an MS in Electrical and Electronics Engineering with thesis in 2025 and a BSc in Computer Engineering with a Mathematics minor and an AI track in 2024 at Koç University, after his first year of undergraduate studies at KAIST. His notable distinctions include an Intel ISEF 4th place Grand Award in Mathematics in 2018. His research interests cover Information Theory, Coding Theory, and Molecular Communication.
\end{IEEEbiography}

    \begin{IEEEbiography}[{\includegraphics[width=1in,height=1.25in,clip,keepaspectratio]{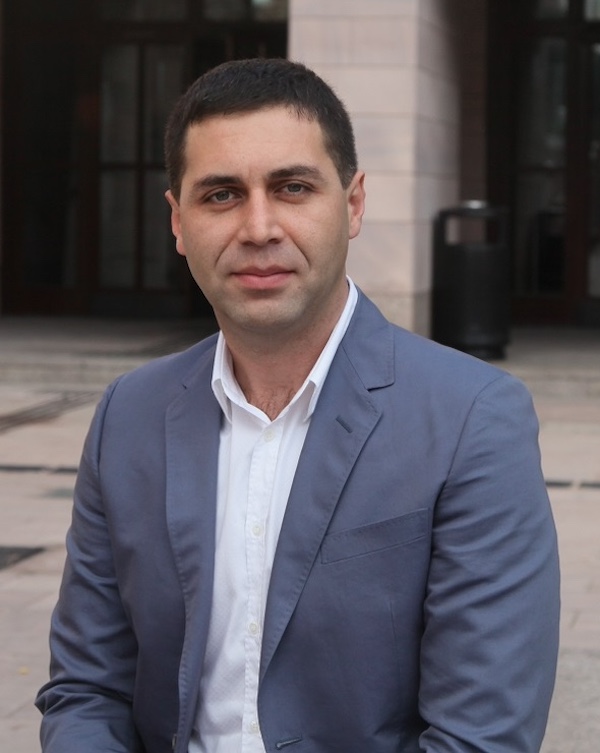}}]{Ozgur B. Akan (Fellow, IEEE)}

 received the PhD from the School of Electrical and Computer Engineering Georgia Institute of Technology Atlanta, in 2004. He is currently the Head of Internet of Everything (IoE) Group, with the Department of Engineering, University of Cambridge, UK and the Director of Centre for neXt-generation Communications (CXC), Koç University, Türkiye. His research interests include wireless, nano, and molecular communications and Internet of Everything.
\end{IEEEbiography}

\end{document}